\documentclass[onecolumn,showpacs,preprintnumbers,amsmath,amssymb,5pt,floatfix]{revtex4}
\usepackage{epsfig,graphics}
\usepackage{dcolumn}
\usepackage{bm}
\usepackage{multirow}
\usepackage{hyperref}
\usepackage{color}
\usepackage{hyperref}
\usepackage{cleveref} 
\usepackage{microtype} 
\usepackage{mathrsfs,amsmath}
\usepackage[version=4]{mhchem}
\usepackage{mathtools}

\usepackage{soul}
\usepackage{chemfig}
\usepackage{xr}

\usepackage{booktabs}

\begin{document}
\newtheorem{thm}{Theorem}[section]
\newtheorem{lemma}[thm]{Lemma}
\newtheorem{prop}[thm]{Proposition}
\newtheorem{rem}[thm]{Remark}
\newtheorem{cor}[thm]{Corollary}

\date{\today}

\title{Thermodynamics of Chemical Reactions in Open Systems Reduced by PEA and QSSA}
\author{
Xinyu Zhang\textsuperscript{2,$\ast$}, 
Haiyang Jia\textsuperscript{1,3,$\ast$}, 
Liangrong Peng\textsuperscript{1,$\dag$}, 
Liu Hong\textsuperscript{2,$\ddag$}
}
\affiliation{
\textsuperscript{1} School of Computer and Data Science, Minjiang University, Fuzhou, Fujian, 350108, P. R. China\\
\textsuperscript{2} School of Mathematics, Sun Yat-Sen University, Guangzhou, Guangdong, 510275, P. R. China\\
\textsuperscript{3} School of Mathematics and Statistics, Fuzhou University, Fuzhou, Fujian, 350108, P. R. China
}

\footnotetext[1]{$\ast$ These authors contributed equally to this work}
\footnotetext[2]{$\dag$ Corresponding author: peng@mju.edu.cn}
\footnotetext[3]{$\ddag$ Corresponding author: hongliu@sysu.edu.cn}


\begin{abstract}
Previous studies have primarily focused on the nonequilibrium thermodynamics of chemical reaction networks (CRNs) occurring in closed systems. In contrast, CRNs in open systems exhibit much richer nonequilibrium phenomena due to sustained matter and energy exchange. Here, we bridge the quantitative relationships between essential thermodynamic quantities --including the steady state, enthalpy, intrinsic Gibbs free energy, and entropy production rate -- in original mass-action equations and their PEA- or QSSA-reduced counterparts for open CRNs. Our analysis demonstrates that the thermodynamic structure, especially the second law of thermodynamics, of the full CRNs may not be preserved in reduced models when algebraic relations are imposed. Specifically, PEA-reduced models lose monotonicity in the intrinsic Gibbs free energy, whereas QSSA retains this property. These theoretical findings are further validated through analytical and numerical studies of two archetypal open systems: the Michaelis-Menten reactions and the phosphorylation-dephosphorylation cycle (PdPC). 
Our results provide a systematic framework for evaluating the fidelity of reduced models. 

\textbf{Keywords:} nonequilibrium thermodynamics, open chemical reactions, complex balance, partial equilibrium approximation, quasi-steady-state approximation
\end{abstract}

\maketitle

\newpage
\section{Introduction}
Chemical reaction networks (CRNs) serve as foundational frameworks for modeling processes spanning chemical synthesis and degradation, energy storage and transduction, and dynamic regulation of biological activation-deactivation cycles \cite{segel1989quasi,alberty2003thermodynamics,wachtel2018thermodynamically}. CRNs are categorized as closed or open based on their interactions with external reservoirs. In contrast to closed systems, open CRNs sustain continuous exchange of specific chemostatted species (e.g. substrates, fuels) with reservoirs, thus enabling the existence of rich nonequilibrium behaviors central to biological energy transduction, cellular regulation, and synthetic biochemical design \cite{feinberg2019foundations,Avanzini2020,qian2021stochastic}.

The thermodynamic framework provides fundamental understanding on the dissipative structure and far-from-equilibrium steady states of open CRNs. A rigorous thermodynamic description requires a quantitative analysis of the coupling among entropy production, energy fluxes, and matter exchange with reservoirs. Although nonequilibrium thermodynamic formalisms for open CRNs governed by elementary mass-action kinetics have been established \cite{feinberg2019foundations,qian2021stochastic,Falasco2025RevModPhys}, persistent challenges remain in characterizing energy dissipation features within coarse-grained models. This difficulty arises primarily from the breakdown of elementary reactions and mass-action laws during the procedure of coarse graining.

Recent advances in the thermodynamics of reduced models have yielded critical insights across multiple methodologies \cite{gorban2005invariant, chiavazzo2007comparison,ge2016mesoscopic,ge2017mathematical, Avanzini2020,yu2021inverse, avanzini2023circuit,shimada2024universal,peng2024thermodynamics}. 
Firstly, thermodynamics provides a valuable framework for developing systematic model reduction approaches in dissipative systems. For instance, Gorban and Karlin \cite{gorban2005invariant} leveraged thermodynamic properties of kinetic equations to derive invariant manifolds, establishing a geometric foundation for reduction. Building on this, Chiavazzo et al. \cite{chiavazzo2007comparison} rigorously compared methods for constructing slow invariant manifolds in chemical kinetics, emphasizing geometric consistency. 
Secondly, emerging methodologies focus on constructing thermodynamically consistent reduced models solely from observable variables, circumventing the knowledge of the full system. For example, Avanzini and colleagues \cite{Avanzini2020,avanzini2023circuit} developed coarse-graining frameworks that embed thermodynamic consistency (e.g. entropy production bounds) directly into reduced models using observable dynamics. This paradigm bridges data-driven reduction techniques with thermodynamic laws. 
Thirdly, simplified models exhibit intrinsic thermodynamic structures that are universal across scales and model-specific details. Ge and Qian \cite{ge2016mesoscopic,ge2017mathematical} demonstrated that macroscopic and mesoscopic descriptions share identical thermodynamic structures under coarse graining. Subsequent studies revealed that energy dissipation rates in coarse-grained CRNs follow an inverse power-law dependence on the number of microstates per coarse-grained macrostate \cite{yu2021inverse}. Furthermore, Shimada et al. \cite{shimada2024universal} derived thermodynamically consistent universal slow dynamics for weakly driven open CRNs, governed by conserved quantities inherited from their closed counterparts. Notably, Peng and Hong \cite{peng2024thermodynamics} compared the thermodynamic quantities in the original closed CRNs and their reduced models via \emph{partial equilibrium approximation} (PEA) or \emph{quasi-steady-state approximation} (QSSA). This study investigated the preservation of the essential thermodynamic structure during the procedure of model reduction, indicating that both PEA and QSSA methods do not necessarily maintain the thermodynamic structure of the full model.  

In this work, we are going to examine a macroscopic CRN constituted by massive particles in a homogeneous and ideal dilute solution. The reactions are assumed to proceed at constant temperature $\mathcal{T}$ and constant pressure $\mathcal{P}$. We suppose there are $N$ species $\{S_1,S_2,\cdots,S_N\}$ participating in the general reversible reactions as follows 
\begin{equation}
\label{r}
\nu _{1i}^+ S_{1} {\text{+}} \nu_{2i}^+ S_{2} {\text{ + }} \cdots {\text{ + }}\nu _{Ni}^+  S_{N}
\xrightleftharpoons[{\kappa_i^-}]{{\kappa_i^+}}
\nu_{1i}^- S_{1} {\text{ + }} \nu_{2i}^- S_{2} {\text{ + }} \cdots {\text{ + }}\nu_{Ni}^-  S_{N},
\end{equation}
where $\nu_{ki}^+\geq 0$ and $\nu_{ki}^-\geq 0$ are stoichiometric coefficients, 
$\kappa _i^+\geq 0$ and $\kappa _i ^-\geq 0$ are the forward and backward rate constants of the $i$'th reaction ($i=1,2,\cdots,M$), respectively. Furthermore, $\nu_{ki}=(\nu_{ki}^- - \nu_{ki}^+)$ denote the elements of the stoichiometric matrix $\bm{\nu}=[(\nu_{ki})]_{N\times M}$. The reversibility of reactions indicates that $\kappa _i^+ >0$ if and only if $\kappa _i^- >0$. The molar concentration of the species $S_k$ is represented as $c_k(t)=[S_k]$, and its vector form $\bm{c}=(c_1, c_2, \cdots, c_N)^{\dag}$ with the superscript $\dag$ denoting the transpose. 

In the companion work for closed CRNs characterized by mass-action laws \cite{peng2024thermodynamics}, the relations between the thermodynamic quantities before and after model reduction by PEA or QSSA were established. However, for chemical reactions occurring in open systems, some species may be exchanged with the environment as a result of controls by external reactions or connections with particle reservoirs. Our current work concerns the applicability of model reduction methods in open CRNs, including PEA and QSSA, from the perspective of thermodynamics. 

The remainder of the paper is organized as follows. In Sec. II, the mass-action equations for open chemical reactions are introduced, along with the PEA and QSSA methods. Section III contains our main results about the nonequilibrium thermodynamics for the reduced models by PEA and QSSA methods separately. The open MM reactions and the phosphorylation-dephosphorylation cycle are studied in detail as concrete examples in Sec. IV and Sect. V respectively. We summarize our results in the last section.

\section{General theory for open chemical reactions}

\subsection{Chemical Mass-action Laws in Macroscopic Scale}
The time evolution of concentrations is characterized through the following general rate equations \cite{van1992},
\begin{equation}
\label{massactioneq_0}
 \frac{{d}}{{dt}}\bm c(t) =  \bm{\nu} \bm R (\bm{c}) + \bm{I}(t), 
\end{equation} 
where $\bm R (\bm{c})$ is the net reaction rate vector, and $\bm I (t)$ is the external current vector. By adopting the decomposition of internal and external-connected species, the concentration vector, external current vector, and stoichiometric matrix are denoted as 
$\bm{c}=(\bm{c}^X,\bm{c}^Y)^{\dag}, \bm{I}=(\bm{0},\bm{I}^Y)^{\dag}, \bm{\nu}=(\bm{\nu}^X,\bm{\nu}^Y)^{\dag}$,
with the sizes of matrices $\bm{c}^X$, $\bm{c}^Y$, $\bm{I}^Y$, $\bm{\nu}^X$ and $\bm{\nu}^Y$ being $N_X \times 1$, $N_Y \times 1$, $N_Y \times 1$, $N_X \times M$ and $N_Y \times M$ respectively, and $N_X+N_Y=N$, the governing equations for the internal species $\bm{c}^X$ and chemostatted species $\bm{c}^Y$, in component form equivalent with \eqref{massactioneq_0}, become  
\begin{subequations}
\label{Conversation law.}
\begin{align}
\label{massactioneq1}
 &\frac{{d}}{{dt}}c_k^X(t) = \sum\limits_{i = 1}^{M} {\nu_{ki}^X} R_i (\bm{c}), \quad k = 1,2,...,N_X, \\ 
\label{massactioneq2}
 &\frac{{d}}{{dt}}c_k^Y(t) = \sum\limits_{i = 1}^{M} {\nu_{ki}^Y} R_i (\bm{c}) + I^Y_k(t), \quad k = N_X+1,N_X+2,...,N,
  \end{align}
\end{subequations}
where $\{I^Y_k(t)\}$ are external currents corresponding to the chemostatted species $\{c^Y_k(t)\}$. If all external currents vanish, ${\bm I}^Y(t) \equiv \bm 0$, the CRN in \eqref{massactioneq_0} becomes a closed system. The net reaction rate function for the $i$'th reaction is $R_i (\bm{c})=R_i^+ (\bm{c}) - R_i^-(\bm{c})$, where the respective forward and backward reaction rates are given by the mass-action law, of a polynomial form as 
\begin{equation}
\label{massactionlaw}
R_i^+(\bm{c}) = \kappa _i ^ +  \prod\limits_{j = 1}^N {c_j ^{\nu _{ji} ^ +  } } ,\quad
R_i^-(\bm{c}) = \kappa _i ^ -  \prod\limits_{j = 1}^N {c_j ^{\nu _{ji} ^ -  } }.
\end{equation}

In what follows, we will introduce three different but related states of the reaction rate equation  \eqref{massactioneq_0}, including the general steady state, complex-balanced steady state and detailed-balanced steady state. It is noticeable that the complex-balanced steady state is usually defined for open CRNs, while the detailed-balanced steady state is defined for closed CRNs, both of which belong to general steady states. 

\textit{\textbf{General steady state}}.
The general steady state $\bm{c}^{ss}=(c^{ss}_1 , c^{ss}_2, \cdots , c^{ss}_N)^{\dag}>0$ of the reaction rate equation in \eqref{massactioneq_0} is introduced as
\begin{equation}
\label{ss}
\bm{\nu R}(\bm{c}^{ss})+\bm{I}^{ss}=\bm 0,
\end{equation}
here $\bm{I}(t)\equiv\bm{I}^{ss}$ denote the stationary external current independent of time.
In the steady state, the total inflow into the species ${S}_k$ ($k=1,2,..., N$) equals to the total outflow from it, contributed by all reactions.

\textit{\textbf{Complex-balanced steady state}}. 
A complex is the combination of all species either on the left-hand or right-hand side of the CRN \eqref{r}. 
The stoichiometric matrix $\bm \nu$ can be decomposed into the product of the respective composition matrix $\bm \Gamma$ and the incidence matrix $\bm \Psi$ as $\bm\nu = \bm\Gamma \bm\Psi$ \cite{Rao2016Nonequilibrium}. The entry \(\Gamma_{ij}\) of the composition matrix \(\bm{\Gamma} = [(\Gamma_{ij})]\) is the stoichiometric number of the species \(S_i\) in the complex \(j\), which encodes the structure of each complex in terms of species. Meanwhile, the entries of the incidence matrix \(\bm{\Psi} = [(\Psi_{jk})]\) are given by 
\begin{equation}
\label{inci mat}
\Psi_{jk} =
\left\{
\begin{aligned}
1 \quad & \text{if } j \text{ is the product complex of the reaction } k, \\
-1 \quad & \text{if } j \text{ is the reactant complex of the reaction } k, \\
0 \quad & \text{otherwise.}
\end{aligned}
\right.
\end{equation}
 
The complexes with the same stoichiometry for the internal species are regrouped into a single set. As a special case, all complexes made solely of chemostatted species are regrouped into the same complex $\mathcal{G}_0$. Therefore, we denote these regrouped complexes as $\mathcal{G}=\{\mathcal{G}_0, \mathcal{G}_1, \cdots, \mathcal{G}_K\}$. Due to the regrouped complexes, the elements of the regrouped incidence matrix are introduced as $\Psi_{lk}^{\mathcal{G}} \equiv \sum_{j \in \mathcal{G}_l}\Psi_{jk}~(l=0,1,\cdots,K;~ k=1,\cdots, M)$. That is, the regrouped incidence matrix is obtained by adding the rows of the incidence matrix \eqref{inci mat} that correspond to the same set to one row. 
The \emph{complex-balanced} steady state $\bm{c}^{cb}=(c^{cb}_1, c^{cb}_2, \cdots , c^{cb}_N)^{\dag}>0$ is defined to have zero net flux in each regrouped complex \cite{Rao2016Nonequilibrium}, 
\begin{equation}
    \label{cb}
    \bm{\Psi}^{\mathcal{G}} \bm{R} (\bm{c}^{cb})=\bm 0.
\end{equation}
Please see Sect. \ref{open Michaelis-Menten reactions} for an illustration of the above definitions. Specifically, the CRNs with zero deficiency are \emph{unconditionally complex-balanced} for any fixed reaction rate constants and any concentrations of chemostatted species. 

\textit{\textbf{Detailed-balanced steady state}}. 
When the external currents vanish, ${\bm I}^Y(t) \equiv \bm 0$, the CRN \eqref{r} and its corresponding evolution equation \eqref{massactioneq_0} degenerate into a closed system. The \emph{detailed balance} guarantees that a closed reaction system, in the steady state, must have zero net flux along each reaction path \cite{Rao2016Nonequilibrium}. Mathematically, the mass-action system in Eq. \eqref{massactioneq_0} is under detailed balance if and only if there exists a positive static state (or thermodynamic equilibrium) $\bm c^e=(c^e_1 , c^e_2, \cdots , c^e_N)^{\dag}>0$ such that 
\begin{equation}
\label{macrocomplex}
\bm R(\bm{c}^e)= \bm 0.
\end{equation} 
Based on above three definitions, it is observed that the detailed balance is a subset of complex balance, and the complex balance is a subset of general steady state. 
In particular, the steady state that violates the detailed-balanced condition is called a nonequilibrium steady state (NESS), which is ubiquitous in open chemical reactions. 

\textit{\textbf{Conservation law}}.
Considering the corresponding closed CRNs, the linearly independent vectors $\left\{\boldsymbol{\ell}_k\right\}_{k=1}^L$ in the left null space of the stoichiometric matrix $\bm{\nu}$, satisfying the condition $\boldsymbol{\ell}_k \cdot  \boldsymbol{\nu} =\bm 0$, represent the conservation laws of Eq. \eqref{massactioneq_0}, where \(L = N-\text{rank}(\boldsymbol{\nu})\) is the number of conservation laws. 

For open CRNs, this definition implies that 
\begin{equation}
    \label{conservlaw_2}
    \frac{d}{dt}\boldsymbol{\ell}_k \cdot \begin{pmatrix}
\bm{c}^X  \\
\bm{c}^Y 
\end{pmatrix} 
= \boldsymbol{\ell}_k \cdot \begin{pmatrix}
\bm{0}  \\
\bm{I}^Y 
\end{pmatrix}.
\end{equation}
If the right-hand side of Eq. \eqref{conservlaw_2} further vanishes, i.e. 
$\boldsymbol{\ell}_k \cdot (\bm{0},\bm{I}^Y)^{\dag}=0$, then \(\boldsymbol{\ell}_k\) represents an unbroken conservation law. Otherwise, if $\boldsymbol{\ell}_k \cdot (\bm{0},\bm{I}^Y)^{\dag}\neq 0$, \(\boldsymbol{\ell}_k\) represents a broken one. Accordingly, the set \(\left\{\boldsymbol{\ell}_k\right\}_{k=1}^L\) is divided into two disjoint subsets, including the set of unbroken conservation laws \(\left\{\boldsymbol{u}_i\right\}_{i=1}^U\) and the set of broken conservation laws \(\left\{\boldsymbol{b}_i\right\}_{i=1}^B\), with $U+B=L$. 



\textit{\textbf{Cycle}}.  
The internal cycles are defined as the independent vectors of the kernel of the stoichiometric matrix, that is $\bm\nu {\bm c}= \bm 0$. Crucially, the existence of cycles precludes thermodynamic equilibrium by sustaining non-vanishing entropy production even at the steady state, driving the system into a nonequilibrium regime.

\subsection{Model Reduction for CRNs}
The dynamical complexity of CRNs grows rapidly with increasing numbers of reactions or interacting species. Classical reduction techniques, including the partial equilibrium approximation (PEA), quasi-steady-state approximation (QSSA), and maximum entropy principle, have been widely employed to mitigate this complexity. Extensive theoretical analyses have established sufficient conditions for the rigorous convergence of PEA- and QSSA-reduced solutions to their full-system counterparts \cite{kokotovic1999singular,yong2012}. In this section, we first outline the basic ideas of PEA and QSSA, then demonstrate their applications to deriving reduced-models from chemical mass-action equations.

\subsubsection{Partial equilibrium approximation}
As to all $M$ reactions, PEA assumes that some reversible ones (saying, the first $W$ reactions) take less time to reach (partial) equilibrium than the rest. These reversible reactions are called fast reactions. After reaching equilibrium, the fast reactions make no contribution to the evolution of concentration for each reactant \cite{gorban2005invariant}. By introducing a small parameter $0<\epsilon\ll 1$ to character the fastness of these reactions, which equals to the ratio of relaxation times of the fast and slow reactions, we can recast the ordinary differential equations (ODEs) in \eqref{massactioneq_0} into  
\begin{equation}
\label{PEA0}
\frac{{dc_k}}{{dt}} = 
\frac{1}{\epsilon}\sum\limits_{i = 1}^W {\nu _{ki}} \overline{R_i}(\bm{c})+
\sum\limits_{i = W + 1}^M {\nu _{ki}} R_i(\bm{c})+ I_k(t), \quad k = 1,2,...,N, 
\end{equation}
where $\overline{R_i}(\bm{c})$ is of the same order for the fast reactions $(i=1,2,\cdots,W)$ as those for the slow reactions $(i=W+1,W+2,\cdots,M)$ by re-scaling the fast ones as $\overline{R_i}(\bm{c}) \equiv {\epsilon}R_i(\bm{c})$ $(i=1,2,\cdots,W)$. 

Using PEA, or equivalently, in the limit of $\epsilon\rightarrow0$, Eq. \eqref{PEA0} degenerates into $W$ algebraic equations and $(N-V)$ ODEs as 
\begin{subequations}
\label{PEA1and2}
\begin{align}
\label{PEA1}
 &R_i ^ +  (\bm{c})=R_i^-  (\bm{c}),\;\;\;i = 1,2,...,W, \\ 
\label{PEA2}
 &\frac{{d}c_k}{{dt}}
 =\sum\limits_{i=W+1}^M \nu _{ki} R_i (\bm{c})+ I_k(t), 
 \quad k = V+1,V+2,...,N. 
  \end{align}
\end{subequations}
Here we assume that the concentrations of the first $V$ ($V\leq W$) species could be explicitly solved from Eq. \eqref{PEA1} as functions of the remaining $(N-V)$ variables. In terms of concreteness, we take the logarithmic transformation on both sides of Eq. \eqref{PEA1}, which leads to
\begin{equation*}
\underbrace{ \sum_{j=1}^V \nu_{ji}\ln c_j }_{\text{the i-th element of}~A\bm{x}}
=\underbrace{ \ln \left(\frac{\kappa _i ^ +}{\kappa _i ^ -} \right)-\sum_{j=V+1}^N \nu_{ji}\ln c_j }_{\text{the i-th element of}~\bm{b}_1}, \quad i = 1,2,...,W.
\end{equation*}
Rewrite it into a matrix form $A^{\dag} \bm{x}=\bm{b}_1$, where the rank of matrix $A=[(\nu_{ji})]_{V\times W}$ is assumed to be $V$, $\bm{x}=(\ln c_1,\cdots, \ln c_V)^{\dag}$ and $\bm{b}_1=(\ln (\kappa _1 ^ +/\kappa _1 ^ -)-\sum_{j=V+1}^N \nu_{j1}\ln c_j,\cdots, \ln (\kappa _W ^ +/\kappa _W ^ -)-\sum_{j=V+1}^N \nu_{jW}\ln c_j)^{\dag}$ are vectors of size $V$ and $W$, respectively. Since $A$ is non-degenerate, the vector $\bm{x}$ could be explicitly solved, $\bm{x}=(AA^{\dag})^{-1}A\bm{b}_1$, which means 
\begin{equation}
\label{PEA_cdef}
(c_1,\cdots,c_V)^{\dag}=
\exp[(AA^{\dag})^{-1}A\bm{b}_1].
\end{equation}

It is well-known that the PEA introduces an initial layer phenomenon, which may lead to inconsistencies in the initial values of the reduced and full models. Mathematically, the presence of the small parameter $\epsilon$ in the highest-order term of Eq. \eqref{PEA0} renders the approximation solution a singular perturbation problem \cite{huang2015partial}. Similar considerations apply to the QSSA, as both methods share analogous mathematical structures in their reduction procedures.

\subsubsection{Quasi-steady-state approximation}
\label{Quasi steady state approximation}
In contrast to rapid balance assumption on reversible reactions by PEA, QSSA assumes that the production and consumption rates of certain species are equal, so that they will stay in a quasi-steady state after a relatively short period of time \cite{gorban2005invariant}. 
Here the concentrations of chemostatted species are assumed to be fixed or driven slowly by the chemostats \cite{Avanzini2020}. Therefore, a part of the internal vector $\bm{c}^X$ are treated as fast species (denoted by $\bm{c}^F(t)$), while the rest part of $\bm{c}^X$ and the whole $\bm{c}^Y$ together are treated as slow species ($\bm{c}^S(t)$). 
Adopting a small parameter $0<\epsilon\ll 1$ to characterize the difference in the relaxation time between the fast species $\bm{c}^F(t)$ and the remaining slow species $\bm{c}^S(t)$, the ODEs in \eqref{massactioneq_0} for $\bm{c}=(\bm{c}^F,\bm{c}^S)$ are rewritten as 
\begin{subequations}
\label{QSSA}
\begin{align}
\frac{{d}}{{dt}}\bm{c}^F &= 
\frac{1}{\epsilon}\bm{\nu}^F 
\bm R(\bm{c}), \label{QSSA1}\\ 
\frac{{d}}{{dt}}\bm{c}^S&= \bm{\nu}^S 
\bm R(\bm{c})+\bm{I}^S(t), \label{QSSA2}
 \end{align}
\end{subequations}
where $\bm{\nu}^F$ and $\bm{\nu}^S$ denote the stoichiometric matrices for fast and slow species respectively based on the partition mentioned above. 

According to QSSA, for every $\bm{c}^S$, there exists a vector $\bm{c}^F(t)=\bm{\widetilde{c}}^F(\bm{c}^S)$ such that $\bm{\nu}^F 
\bm R(\bm{\widetilde{c}}^F(\bm{c}^S),\bm{c}^S)=\bm 0$. Therefore, the vector $R(\bm{\widetilde{c}}^F(\bm{c}^S),\bm{c}^S)$ 
is a linear combination of the right null eigenvectors of $\bm{\nu}^F$, that is $R(\bm{\widetilde{c}}^F(\bm{c}^S),\bm{c}^S)=\sum_{\gamma} \bm{\phi}_{\gamma}{\psi}_{\gamma}(\bm{c}^S)$, where $\bm{\nu}^F \bm{\phi}_{\gamma}=\bm{0}$ and $\{{\psi}_{\gamma}(\bm{c}^S)\}$ denotes the coefficients. By using the topological concepts of internal cycle, $\bm{\nu}^F \bm{\phi}_l=\bm{0}$ and $\bm{\nu}^S \bm{\phi}_l=\bm{0}$, and pseudo-emergent cycle, $\bm{\nu}^F \bm{\phi}_e= \bm{0}$ and $\bm{\nu}^S \bm{\phi}_e\neq \bm{0}$, we have the relation $\bm{\nu}^S 
\bm R(\bm{c})=\sum_{\gamma}\bm{\nu}^S 
\bm{\phi}_{\gamma}{\psi}_{\gamma}(\bm{c}^S)=\sum_{e}\bm{\nu}^S 
\bm{\phi}_{e}{\psi}_{e}(\bm{c}^S)$. 
The original mass-action equations in \eqref{QSSA} are thus reduced to 
\begin{subequations}
\label{reQSSA}
\begin{align}
\frac{{d}}{{dt}}\bm{c}^S&=\bm{\widetilde \nu}^S\bm{\psi}(\bm{c}^S)+\bm{I}^S(t), \\
\bm{c}^F&=\bm{\widetilde{c}}^F(\bm{c}^S), 
\end{align}
\end{subequations}
where $\bm{\widetilde \nu}^S$ and $\bm{\psi}$ represent the effective stoichiometric matrix and effective current vector, whose $e$'th column $\bm{\widetilde \nu}^S_e =\bm{\nu}^S
\bm{\phi}_{e}$ and $e$'th element ${\psi}_e$. 



\section{nonequilibrium thermodynamics for reduced models of open CRNs}
\label{ther reduced}

\subsection{Nonequilibrium Thermodynamics of Full CRNs}
\label{noneq ther}
Chemical thermodynamics furnishes the fundamental principles of energy and forces within the realm of chemical species and their changes \cite{qian2021stochastic}. Thanks to the rapid development of stochastic thermodynamics and steady-state thermodynamics in recent years \cite{Rao2016Nonequilibrium,ge2016mesoscopic,ge2017mathematical}, the chemical reactions has rich connotations of thermodynamics. To be concrete, the essential thermodynamic properties of open CRNs not only include the fundamental laws of thermodynamics, but also include the (nonequilibrium) stationary state, entropy, enthalpy, free energy, entropy flow, entropy production, free energy dissipation, and etc. In the present work, we focus on the first and second laws of thermodynamics, enthalpy, entropy and free energy functions, as well as their time derivatives, and their decompositions. 

\subsubsection{The first law of thermodynamics}
In a general system, \textit{\textbf{the first law of thermodynamics}} states that the change in the internal energy $\Bar{\Bar{U}}$ of the system is equal to the heat $\Bar{\Bar{Q}}$ and work done $\Bar{\Bar{W}}$ on the system: 
\begin{equation}
\label{general 1st law}
   \Delta \Bar{\Bar{U}} = \Bar{\Bar{Q}} + \Bar{\Bar{W}},
\end{equation}
here the extensive variable $\Bar{\Bar{U}}$ is used to denote the overall internal energy of the system. 

For a system at constant pressure, the work is given by $\Bar{\Bar{W}}=-\mathcal{P} \Delta V$, here the minus sign guarantees that the work done on the system is positive ($\Bar{\Bar{W}} > 0$) when the volume is compressed  ($\Delta V < 0$). On the other hand, the heat transferred at constant pressure is described by the enthalpy change, $\Bar{\Bar{Q}}=\Delta H$. Substituting these terms into the general first law \eqref{general 1st law}, we have 
\begin{equation*}
   \Delta \Bar{\Bar{U}} 
   =\Delta H -\mathcal{P} \Delta V 
   =\Delta(H-\mathcal{P}V),
\end{equation*}
which deduces that the internal energy $\Bar{\Bar{U}}$ and the enthalpy $\Bar{\Bar{H}}$ are related via 
\begin{equation}
\label{general 1st law_2}
    \Bar{\Bar{U}} = \Bar{\Bar{H}} - \mathcal{P}V, 
\end{equation}
up to a constant. 

Specifically, the volume of solution $V$ is assumed to be constant since it is overwhelmingly dominated by the solvent. Once divided by the volume $V$, Eq. \eqref{general 1st law_2} becomes $\Bar{\Bar{U}}/V = \Bar{\Bar{H}}/V - \mathcal{P}$. Hereinafter we focus on the intensive variables, such as the internal energy density $U\equiv \Bar{\Bar{U}}/V$ and enthalpy density $H=\Bar{\Bar{H}}/V$, and neglect the word density without ambiguity. Therefore, the internal energy equals the enthalpy up to a constant, $U=H-\mathcal{P}$. 

\textit{\textbf{Enthalpy}}. 
The enthalpy for open CRNs is defined as 
\begin{equation}
\label{enthalpy}
H(t)=\bm{c} \cdot \bm{h}^{\circ} + H_0, 
\end{equation}
where $\bm{h}^{\circ}$ denotes {the standard-state enthalpy of formation}, the dot $\cdot$ denotes the scalar product, $H_0$ is a constant. 
Correspondingly, the change rate of enthalpy \eqref{enthalpy} becomes 
\begin{equation}
\label{dHdt}
    \frac{dH}{dt}={\bm h^{\circ}} \cdot {\bm \nu} {\bm R (\bm{c})} + {\bm h^{\circ}} \cdot \bm{I}(t), 
\end{equation}
where the first term ${\bm h^{\circ}} \cdot {\bm \nu} {\bm R (\bm{c})}$ is identified as the heat flow rate, while the second term ${\bm h^{\circ}} \cdot \bm{I}(t)$ is the enthalpy exchange rate. 
Notice that, Eq. \eqref{dHdt} is considered to be the nonequilibrium analogy of the first law of thermodynamics for closed CRNs \cite{Rao2016Nonequilibrium}.

\subsubsection{The second law of thermodynamics}

The \textit{\textbf{second law of thermodynamics}} mainly deals with the mathematical properties of the entropy function \(Ent(t)\). For the CRN \eqref{r}, the entropy is chosen as 
\begin{equation}
\label{entropy}
Ent(t)=  - \mathcal{R}\bm{c} \cdot (\ln {\bm c} - {\bm 1})+\bm{c} \cdot {\bm s}^{\circ} + Ent_0,
\end{equation} 
where $\mathcal{R}$ is the gas constant, ${\bm 1}$ is a vector with all components of $1$, ${\bm s}^{\circ}$ is the standard entropy of formation \cite{grmela2012fluctuations,ge2013dissipation, Rao2016Nonequilibrium}, $(\mathcal{R}\bm{c} \cdot {\bm 1}+Ent_0)$ denotes the entropic contribution of the solvent. For notation simplicity, the constant $Ent_0$ is set to be $0$ in what follows.

Combining the entropy with the chemical reaction rate equation in Eq. \eqref{massactioneq_0}, we can derive the following \textit{\textbf{entropy balance equation}},
\begin{subequations}
\begin{align}
&\frac{d}{{dt}}Ent(t) = J^f(t)  + epr(t),\\
&J^f(t)= ({\bm s^{\circ}} - \mathcal{R}\ln {\bm c}) \cdot \bm{I}(t) + {\bm s^{\circ}} \cdot {\bm \nu} {\bm R (\bm{c})}- \mathcal{R} \bm{R} (\bm{c}) \cdot \ln \frac{\bm \kappa^+}{\bm \kappa^-}, \\
&epr(t)=
\mathcal{R} [\bm R^+  (\bm{c}) - \bm R^-  (\bm{c})] \cdot \ln \frac{{\bm R^+ (\bm{c})}}{{\bm R^- (\bm{c})}} \geq0,
\end{align}
\end{subequations}
where \(J^f(t) \) and \(epr(t)\) denote the entropy flux and entropy production rate respectively. The entropy flux $J^f(t)$ and entropy production rate $epr(t)$ are separated according to their different physical origins and mathematical properties. The entropy flux $J^f(t)$ represents the entropy flow into the system from the environment due to the exchange of matter and energy, while the entropy production rate $epr(t)$ gives the entropy change in the system induced by irreversible processes, per second. According to the second law of thermodynamics, the entropy production rate $epr(t)$ is always non-negative $epr(t) \geq 0$, and $epr(t)=0$ if and only if the system is at equilibrium. Concerning with CRNs, we have $epr(t)=0$ if and only if $\bm R^ +  (\bm{c}^e) =\bm R^-  (\bm{c}^e)$.

\textit{\textbf{Free energy}}. 
With respect to the entropy and enthalpy, the Gibbs free energy function reads 
\begin{equation}
G(t)=H(t)-\mathcal{T}\cdot Ent(t)
=\bm{c} \cdot ({\bm h}^{\circ} - \mathcal{T} {\bm s}^{\circ} + \mathcal{RT} \ln {\bm c}) - \mathcal{RT}\bm{c} \cdot {\bm 1} + (H_0 - \mathcal{T}\cdot Ent_0).
\end{equation}
In terms of the chemical potentials $\bm{\mu}={\bm \mu}^{\circ} + \mathcal{RT} \ln {\bm c}$, we can rewrite the Gibbs free energy as $G(t)=\bm{c} \cdot \bm{\mu}-\mathcal{RT}\bm{c} \cdot {\bm 1} + G_0$ with $G_0=H_0 - \mathcal{T}\cdot Ent_0$. 
In the ideal dilute solution, the standard chemical potential $\bm{\mu}^{\circ}={\bm h}^{\circ} - \mathcal{T} {\bm s}^{\circ}$ is constant, and the expression $\mu_k({c_k})=\mu_k^{\circ} + \mathcal{RT} \ln {c_k}$ indicates that a simplified relation between the chemical potential and the activity is adopted. 

It is a pity that the monotonicity of free energy function with respect to time $t$ for closed CRNs does not hold any more in open systems, due to the presence of external fluxes $\bm{I}^Y(t)$. Therefore, we introduce an alternative quantity -- the intrinsic Gibbs free energy $\mathcal{G}(t)$ of open systems as
\begin{equation}
    \mathcal{G}(t)=G(t)-\int_0^{t} \bm{\mu}^Y(\bm{c}(s)) \cdot \bm{I}^Y(s)ds,
\end{equation}
where $\bm{\mu}^Y(\bm{c})=[\mu^Y_k(\bm{c})]_{k=N_X+1}^N$ is the chemical potential of chemostatted species $\bm{c}^Y$, and $\bm{I}^Y$ is the external current vector of the open system. 
More importantly, its time derivative reads 
\begin{equation}
\frac{d\mathcal{G}}{dt}
=\bm{\mu}(\bm{c}) \cdot \left(\bm{\nu R}(\bm{c})+\bm{I}(t)\right) - \bm{\mu}^Y(\bm{c}) \cdot \bm{I}^Y
=\bm{\mu}(\bm{c}) \cdot \bm{\nu R}(\bm{c}).
\end{equation}
When the system further satisfies the condition of local detailed balance:
\begin{equation}
    \frac{\bm{R}^+({\bm{c}})}{\bm{R}^-({\bm{c}})}=\exp{\left(-\frac{\bm{\mu}({\bm{c}}) \cdot \bm{\nu}}{\mathcal{RT}}\right)},
    \label{ldb}
\end{equation}
we can show that the intrinsic free energy is non-increasing, since $-d\mathcal{G}/dt=\mathcal{T} epr(t) \geq 0$, and $d\mathcal{G}/dt=0$ if and only if $\bm{R}^+({\bm{c}^e}) = \bm{R}^-({\bm{c}^e})$. {Based on the relations among the above thermodynamic quantities, it is observed that $d(H-\mathcal{T}\cdot Ent)/dt=-\mathcal{T} epr(t) + \bm \mu^Y(\bm c) \cdot \bm I^Y(t)$, and therefore $dH/dt=\mathcal{T}J^f(t) + \bm \mu^Y(\bm c) \cdot \bm I^Y(t)$ for isothermal CRNs.}

\textit{\textbf{Relative entropy}}. 
Supposing open CRNs further satisfy the condition of complex balance, we proceed to introduce the relative entropy function $F(t)$ based on the complex-balanced steady state of the internal species $\bm{c}_{X}^{cb}$ as 
\begin{equation}
\label{free}
F(t)
={\mathcal{RT}} \left( {\bm c^X} \cdot \ln\frac{\bm c^X}{\bm{c}_{X}^{cb}} - {\bm c^X} \cdot \bm 1 + \bm{c}_{X}^{cb} \cdot \bm 1 \right)
\ge 0. 
\end{equation}
Note in the above definition, only the concentrations of internal species are involved, which leads to a dramatic distinction from the relative entropy function for close CRNs. In the latter, concentrations of all species, not a part of them, come into play. It is straightforward to show that $F(t)$ is nonnegative and becomes zero if and only if the system reaches the complex-balanced steady state $\bm{c}^{X}=\bm{c}_{X}^{cb}$. 

Correspondingly, the free energy dissipation rate, $f_d(t)$, is defined as the negative value of the time change rate of $F(t)$, 
\begin{align}
f_d(t)  
&\equiv -\frac{dF(t)}{dt}
={\mathcal{RT}} \bm R(\bm{c}) \cdot \ln \frac{{\bm R^+ (\bm{c}) \bm R^-(\bm{c}^{cb}) }}{{\bm R^-  (\bm{c}) \bm R^+  (\bm{c}^{cb})}} \ge 0,
\end{align}
which is also non-negative, reflecting that the relative entropy is non-increasing along the trajectories of the mass-action equations (See \textcolor{blue}{SI}
for details). 
The free energy dissipation rate becomes zero, $f_d(t)=0$, if and only if when the steady state is reached. Thus, $F(t)$ serves as a Lyapunov function for the system of chemical mass-action equations, and the steady state is locally asymptotically stable.

The difference between the rates of entropy production and free energy dissipation scaled by temperature $\mathcal{T}$ gives the housekeeping heat  \cite{ge2016mesoscopic,Rao2016Nonequilibrium,fang2019nonequilibrium}, which describes the heat exchange between the system and its environment in order to maintain a steady state: 
\begin{equation}
Q_{hk}(t)\equiv epr(t) - \frac{1}{\mathcal{T}}f_d(t)  
={\mathcal{R}}\bm R(\bm{c}) \cdot \ln \frac{{\bm R^+ (\bm{c}^{cb} )}}{{\bm R^- (\bm{c}^{cb} )}} 
\ge 0,
\end{equation}
here $Q_{hk}=0$ if and only if the system satisfies the condition of detailed balance, that is, $\bm{c}^{cb}\equiv \bm c^e$ (See \textcolor{blue}{SI}
for details). Note that $Q_{hk}$ and $f_d$ are also referred to as the adiabatic and non-adiabatic entropy production rates, respectively, in some references \cite{ge2016mesoscopic,Rao2016Nonequilibrium,fang2019nonequilibrium}.

\begin{rem}
    For closed CRNs, the external current $\bm I(t) \equiv 0$. The local detailed balance condition guarantees that the steady state must be detailed-balanced, and therefore be a thermodynamic equilibrium. In this case, the relative entropy $F(t)$ and the Gibbs free energy function $G(t)$ satisfy that $G(t)=\mathcal{R T} \cdot F(t)+G_{eq}$, here $G_{eq}$ denotes the Gibbs free energy at equilibrium. Meanwhile, the housekeeping heat $Q_{hk}\equiv 0$. 
\end{rem}

The non-negativity of entropy production rate, free energy dissipation rate, housekeeping heat, $epr, f_d, Q_{hk} \geq 0$, and the monotonicity of intrinsic Gibbs free energy, $d_t\mathcal{G} \leq 0$, serve as different faces of the second law of thermodynamics in open CRNs. We mention that the above formulation holds for the full system before reduction.  
In the next part, we will look into the thermodynamic properties of the reduced system after applying PEA or QSSA, and especially focus on the correspondence between the reduced model and the original full model from the thermodynamic perspective.

\subsection{Thermodynamics of Open CRNs Reduced by PEA}
\label{PEA thermodynamics}
Under PEA, $(c_1,\cdots,c_V)^{\dag}=\mathcal{C}\left(c_{V+1},\cdots,c_N\right)$, 
here $\mathcal{C}$ is the exponential function defined in Eq. \eqref{PEA_cdef}. We denote the solution to the PEA-reduced dynamics as $\overline{\bm{c}}=(\mathcal{C}\left(c_{V+1},\cdots,c_N\right),c_{V+1},\cdots,c_N)^{\dag}$. 
In the following, the overline \(\overline{\bm{c}}\) is used to denote the PEA-reduced results. 

\subsubsection{The first law of thermodynamics}
\textit{\textbf{Enthalpy}}.
For the PEA-reduced model of CRNs, the enthalpy becomes 
\begin{equation}
\label{enthalpy-PEA}
    \overline{H}(t)=\overline{\bm{c}} \cdot \bm{h}^{\circ}+H_0.
\end{equation} 
Correspondingly, the change rate of enthalpy becomes 
\begin{equation}
\label{dHdt_PEA}
    \frac{d\overline{H}}{dt}
    =({h}^{\circ}_1,\cdots, {h}^{\circ}_V)\cdot 
    \frac{d}{dt}\exp[(AA^{\dag})^{-1}A\bm{b}_1] 
    +\sum\limits_{k = V + 1}^N \sum\limits_{i = W + 1}^M {h}^{\circ}_k \nu _{ki} R_i (\overline{\bm{c}}) + \sum\limits_{k = V + 1}^N{h}^{\circ}_k I_k(t). 
\end{equation}
The above formula can be recognized as the first law of thermodynamics for the PEA-reduced model.

\subsubsection{The second law of thermodynamics}
For the PEA-reduced model, the entropy function becomes 
\begin{equation}
\label{entropy-PEA}
\overline{Ent}(t)=  -\mathcal{R}
\overline{\bm{c}} \cdot 
\ln {\overline{\bm{c}}}  +\mathcal{R}\overline{\bm{c}} \cdot {\bm {1}}
+{\overline{\bm{c}}} \cdot {\bm s}^{\circ}.
\end{equation}
Taking time derivative of the entropy function and substituting the dynamics in \eqref{PEA1and2}, we have:
\begin{subequations}
\begin{align}
\frac{d}{{dt}}\overline{Ent}(t) = &\overline{J^f}(t)  + \overline{epr}(t),\\
\overline{J^f}(t)
=&\left[({s}^{\circ}_{1}, \cdots, {s}^{\circ}_{V})^{\dag} - \mathcal{R} (AA^{\dag})^{-1}A\bm{b}_1 \right] \cdot
\frac{d}{dt} \exp[(AA^{\dag})^{-1}A\bm{b}_1] \nonumber\\
&+\mathcal{R} \sum\limits_{i = W + 1}^M  R_i( \overline{\bm{c}} )(\nu_{1i}, \cdots,  \nu_{Vi})^{\dag} \cdot (AA^{\dag})^{-1}A\bm{b}_1 \nonumber\\
&-\sum\limits_{i = W+1}^M 
\left(\mathcal{R} \ln \frac{\kappa_i^+ }{\kappa_i^- } - \sum_{k=V+1}^N s_k^{\circ} \nu_{ki}\right) R_i(\overline{\bm{c}})
+ \sum\limits_{k= V+1}^N (s_k^{\circ} - \mathcal{R} \ln \overline{c_k}) I_{k},\\
\overline{epr}(t)=&
\mathcal{R} \sum\limits_{i=W+1}^M [R^+_i  (\overline{\bm{c}}) - R^-_i  (\overline{\bm{c}})] \ln \frac{{R^+_i (\overline{\bm{c}})}}{{R^-_i (\overline{\bm{c}})}} \geq0. 
\end{align}
\end{subequations}

\textit{\textbf{Free energy}}. 
The intrinsic Gibbs free energy $\overline{\mathcal{G}}(t)$ for the PEA-reduced model reads 
\begin{equation}
\label{intrinsic Gibbs-PEA}
\overline{\mathcal{G}}(t)= \bm{\mu}\left(\overline{\bm{c}}\right) \cdot \overline{\bm{c}} - \mathcal{RT}\overline{\bm{c}} \cdot \bm{1} 
- \int_0^{t}\bm{\mu}\left(\overline{\bm{c}}(s)\right) \cdot \bm{I}(s) ds. 
\end{equation}
Its time derivative is consequently derived as 
\begin{equation}
    \frac{d\overline{\mathcal{G}}}{dt}
    = \bm{\mu}\left(\overline{\bm{c}}\right) \cdot \frac{d \overline{\bm{c}}}{dt} - \bm{\mu}\left(\overline{\bm{c}}\right) \cdot \bm{I}(t).
\end{equation}
It is noticeable that, both the relations  ${d\overline{\mathcal{G}}}/{dt}
    =-\mathcal{T} \overline{epr}(t)$ and ${d\overline{\mathcal{G}}}/{dt}
    \leq 0$ are broken for the PEA-reduced models of CRNs, due to the adoption of algebraic equations.

\textit{\textbf{Relative entropy}}. 
Denote the concentration of internal species for the PEA-reduced dynamics as $\overline{\bm{c}^{X}}=(\mathcal{C}\left(c_{V+1},\cdots,c_N\right),c_{V+1},\cdots,c_{N_X})$.Thus $\overline{\bm{c}^{X}}$ is the internal part of $\overline{\bm{c}}$. 
We proceed to introduce the relative entropy function $\overline{F}(t)$ based on the steady state of the PEA-reduced model $\overline{\bm{c}_{X}^{ss}}$ as
\begin{equation}
\label{relative-PEA}
\overline{F}(t)
={\mathcal{RT}} \left( \overline{{\bm c^X}} \cdot \ln\frac{\overline{{\bm c^X}}}{\overline{\bm{c}_{X}^{ss}}} - \overline{{\bm c^X}} \cdot \bm 1 + \overline{\bm{c}_{X}^{ss}} \cdot \bm 1 \right)
\ge 0. 
\end{equation}
Here, $\overline{F}(t)$ becomes zero if and only if the system reaches the steady state, $\overline{\bm{c}^{X}}=\overline{\bm{c}_{X}^{ss}}$. The free energy dissipation rate, $\overline{f_d}(t)$, is defined as the negative value of the time change rate of $\overline{F}(t)$,
\begin{align}
\overline{f_d}(t)  
&\equiv -\frac{d\overline{F}(t)}{dt}
=-{\mathcal{RT}} (AA^{\dag})^{-1}A(\bm{b}_1 - \bm{b}_1^{ss}) \cdot \frac{d}{dt}\exp[(AA^{\dag})^{-1}A\bm{b}_1] 
 -{\mathcal{RT}} \sum\limits_{i = W + 1}^M R_i (\overline{\bm{c}}) \ln\prod_{k = V + 1}^{N_X}  \left(\frac{\overline{c_k}}{\overline{c_k^{ss}}}\right)^{\nu_{ki}},
\end{align}
where $\bm{b}_1^{ss} \equiv (\ln (\kappa _1 ^ +/\kappa _1 ^ -)-\sum_{j=V+1}^N \nu_{j1}\ln \overline{c_j^{ss}},\cdots, \ln (\kappa _W ^ +/\kappa _W ^ -)-\sum_{j=V+1}^N \nu_{jW}\ln \overline{c_j^{ss}})^{\dag}$ is the vector $\bm{b}_1$ in the steady state for the PEA-reduced model. 
The loss of non-negativity in the free energy dissipation rate in PEA could be attributed to the breakdown of the mass-action law during the adoption of algebraic relations as well as the breakdown of complex-balanced condition.

The housekeeping heat  is deduced by subtracting the free energy dissipation rate from the entropy production rate: 
\begin{align}
\overline{Q_{hk}}(t) 
=&\overline{epr}(t) - \frac{1}{\mathcal{T}}\overline{f_d}(t)\nonumber\\
=&\mathcal{R} \sum\limits_{i=W+1}^M R_i  (\overline{\bm{c}}) 
\ln \left[ \frac{{R^+_i (\overline{\bm{c}})}}{{R^-_i (\overline{\bm{c}})}} \prod_{k = V + 1}^{N_X}  (\frac{\overline{c_k}}{\overline{c_k^{ss}}})^{\nu_{ki}} \right]
+ \mathcal{R} (AA^{\dag})^{-1}A(\bm{b}_1 - \bm{b}_1^{ss}) \cdot \frac{d}{dt}\exp[(AA^{\dag})^{-1}A\bm{b}_1].
\end{align}
Again, the non-negativity of the free energy dissipation rate is no longer guaranteed for PEA-reduced models. 

\subsection{Thermodynamics of Open CRNs Reduced by QSSA}
\label{QSSA thermodynamics}
Let's proceed to discuss the thermodynamics of QSSA-reduced models. Denote the solution to the QSSA-reduced model in Eq. \eqref{reQSSA} as $\widetilde{\bm c}=(\widetilde{\bm c}^F,\widetilde{\bm c}^S)^{\dag}$. In the following, the tilde $\widetilde{\bm c}$
is used to denote the QSSA-reduced results.

\subsubsection{The first law of thermodynamics of QSSA}
\textit{\textbf{Enthalpy}}.
For the QSSA-reduced model of CRNs, the enthalpy becomes 
\begin{equation}
\label{enthalpy-QSSA}
    \widetilde{H}(t)=\widetilde{\bm{c}} \cdot \bm{h}^{\circ}+H_0.
\end{equation} 
Correspondingly, the change rate of enthalpy becomes 
\begin{equation}
\label{dHdt_QSSA}
    \frac{d\widetilde{H}}{dt}
    ={\bm h}^{\circ}_S \cdot [\bm{\nu}^S \bm R(\widetilde{\bm{c}})+\bm{I}^S(t)] 
    +{\bm h}^{\circ}_F \cdot \frac{d\bm{\widetilde{c}}^F}{dt} ,
\end{equation}
where ${\bm h}^{\circ}_S$ and ${\bm h}^{\circ}_F$ are the standard enthalpies of formation for slow and fast species, ${\bm c}^{S}$ and ${\bm c}^{F}$, respectively. 

\subsubsection{The second law of thermodynamics of QSSA} 
The entropy for the QSSA-reduced model is 
\begin{equation}
\label{entropy-QSSA}
\widetilde{Ent}(t)=  -\mathcal{R}\widetilde{\bm{c}} \cdot\ln {\widetilde{\bm{c}}}  
+\mathcal{R}\widetilde{\bm{c}} \cdot {\bm {1}}
+\widetilde{\bm{c}} \cdot {\bm s}^{\circ}.
\end{equation}
Taking time derivative of the entropy function and substituting the reduced governing equation in \eqref{reQSSA}, we have 
\begin{subequations}
\begin{align}
&\frac{d}{{dt}}\widetilde{Ent}(t) 
= \widetilde{J^f}(t)  + \widetilde{epr}(t),\\
&\widetilde{J^f}(t)
= (\bm {s}_S^{\circ} - \mathcal{R}\ln \widetilde{{\bm {c}}}^S) \cdot \bm{I}^S(t) 
+ {\bm{s}_S^{\circ}} \cdot {\bm \nu}^S \bm R (\widetilde{\bm{c}})+ (\bm {s}_F^{\circ} - \mathcal{R}\ln \widetilde{{\bm {c}}}^F) \cdot \frac{d\bm{\widetilde{c}}^F}{dt} 
+\mathcal{R} \bm R (\widetilde{\bm{c}})  \cdot ((\bm {\nu}^F)^{\dag} \ln {\widetilde{\bm {c}}^F} - \ln \frac{\bm \kappa^+}{\bm \kappa^-}), \\
&\widetilde{epr}(t)
=\mathcal{R} [\bm R^+ (\widetilde{\bm{c}}) - \bm R^- (\widetilde{\bm{c}})] \cdot \ln \frac{\bm R^+ (\widetilde{\bm{c}})}{\bm R^- (\widetilde{\bm{c}})} 
\geq0, 
\end{align}
\end{subequations}
where ${\bm s}^{\circ}_S$ and ${\bm s}^{\circ}_F$ are the standard entropies of formation for slow and fast species respectively. The non-negativity of the entropy production rate is a manifestion of the second law of thermodynamics in CRNs.

\textit{\textbf{Free energy}}. 
The intrinsic Gibbs free energy $\widetilde{\mathcal{G}}(t)$  for the QSSA-reduced model becomes 
\begin{equation}
\label{intrinsic-QSSA}
    \widetilde{\mathcal{G}}(t)=\bm{\mu}^S \cdot \widetilde{\bm{c}}^S - \mathcal{RT} \widetilde{\bm{c}}^S \cdot \bm{1}-\int_0^{t}\bm{\mu}^{{S}}(\widetilde{\bm{c}}(s))\cdot\bm{I}^S(s) ds.
\end{equation} 
Consequently, we obtain  
\begin{equation}
    \frac{d\widetilde{\mathcal{G}}}{dt}
    =\bm{\mu}^S \cdot [\bm{\nu}^S\bm R(\widetilde{\bm{c}})+\bm{I}^S(t)] - \bm{\mu}^{S} \cdot \bm{I}^S(t)
    =\bm{\mu}^S \cdot \bm{\nu}^S \bm R(\widetilde{\bm{c}}).
\end{equation} 

Recalling the concepts of internal cycles and pseudo-emergent cycles, $\bm{\nu}^F \bm{\phi}_{\gamma}= \bm{0}$, we have $\bm{\mu}^F \cdot \bm{\nu}^F \sum_{\gamma} \bm{\phi}_{\gamma} {\psi}_{\gamma}(\widetilde{\bm{c}}^S)=\bm{\mu}^F \cdot \sum_{\gamma} (\bm{\nu}^F \bm{\phi}_{\gamma}) {\psi}_{\gamma}(\widetilde{\bm{c}}^S)=0$. 
Substituting the relation $\bm R(\widetilde{\bm{c}})=\sum_{\gamma} \bm{\phi}_{\gamma} {\psi}_{\gamma}(\widetilde{\bm{c}}^S)$ into the above time derivative $d\widetilde{\mathcal{G}}/dt$, we have 
\begin{align}
    \frac{d\widetilde{\mathcal{G}}}{dt}
    &=\bm{\mu}^S \cdot \bm{\nu}^S \sum_{\gamma} \bm{\phi}_{\gamma} {\psi}_{\gamma}(\widetilde{\bm{c}}^S)
     =(\bm{\mu}^S \cdot \bm{\nu}^S+\bm{\mu}^F \cdot \bm{\nu}^F) \sum_{\gamma} \bm{\phi}_{\gamma} {\psi}_{\gamma}(\widetilde{\bm{c}}^S) \nonumber \\
    &=\bm{\mu} \cdot \bm{\nu} \sum_{\gamma} \bm{\phi}_{\gamma} {\psi}_{\gamma}(\widetilde{\bm{c}}^S)
    =-\mathcal{RT} ({{\bm R^+(\widetilde{\bm c})} - {\bm R^-}(\widetilde{\bm c})})\ln \frac{{\bm R^+}(\widetilde{\bm c})}{{\bm R^-}(\widetilde{\bm c})} 
    \leq 0.
\end{align}
In the third equality, we used the partitioning of matrices $\bm{\mu}=(\bm{\mu}^F,\bm{\mu}^S)^{\dag}$ and $\bm{\nu}=(\bm{\nu}^F,\bm{\nu}^S)^{\dag}$; while in the last equality, we used the local detailed balance condition that modified from Eq. \eqref{ldb} under QSSA as ${\bm R^+}(\widetilde{\bm c})/{\bm R^-}(\widetilde{\bm c})=\exp{\left(-\frac{\bm{\nu} \cdot \bm\mu(\widetilde{\bm c}) }{\mathcal{RT}}\right)}$. 

\textit{\textbf{Relative entropy}}. 
At the same time, the relative entropy between concentrations of slow species $\widetilde{\bm c}^S$ and those at the steady state $\widetilde{\bm c_S^{ss}}$ remains non-negative, 
\begin{equation}
    \label{relative-QSSA}
\widetilde{F}(t)
={\mathcal{RT}} \left( \widetilde{{\bm c^S}} \cdot \ln\frac{\widetilde{\bm c^S}}{\widetilde{\bm{c}_{S}^{ss}}} - \widetilde{{\bm c^S}} \cdot \bm 1 + \widetilde{\bm{c}_{S}^{ss}} \cdot \bm 1 \right) \geq 0.
\end{equation} 
Nevertheless, the free energy dissipation rate reads 
\begin{eqnarray}
\widetilde{f_d}(t)
&=&-\mathcal{RT} \ln\frac{\widetilde{\bm c^S}}{\widetilde{\bm{c}_{S}^{ss}}} \cdot \frac{d \widetilde{\bm c^S}}{dt} 
=-\mathcal{RT} {\bm R}(\widetilde{\bm c}) \cdot \ln \left(\frac{\widetilde{\bm c^S}}{\widetilde{\bm{c}_{S}^{ss}}} \right)^{\bm \nu^S}
-\mathcal{RT} \bm I^S(t) \cdot \ln \frac{\widetilde{\bm c^S}}{\widetilde{\bm{c}_{S}^{ss}}},
\end{eqnarray}
whose sign is undetermined. 
Subtracting the free energy dissipation rate from the entropy production rate, $\widetilde{epr}(t) - \widetilde{f_d}(t)/{\mathcal{T}}$, we obtain the housekeeping heat  as 
\begin{align}
\label{Qhk QSSA}
\widetilde{Q_{hk}}(t)
=&\mathcal{R} {\bm R}(\widetilde{\bm c}) \cdot \ln \left[\frac{\bm R^+ (\widetilde{\bm{c}})}{\bm R^- (\widetilde{\bm{c}})} \left(\frac{\widetilde{\bm c^S}}{\widetilde{\bm{c}_{S}^{ss}}}\right)^{\bm \nu^S} \right]
+\mathcal{R} \bm I^S(t) \cdot \ln \frac{\widetilde{\bm c^S}}{\widetilde{\bm{c}_{S}^{ss}}}.
\end{align}
The non-negativity of $\widetilde{Q_{hk}}(t) $ is not guaranteed too. 

In summary, (1) No matter for PEA or QSSA-reduced models, the steady-state is no longer consistent with the original one. 
(2) The loss of non-negativity in the free energy dissipation rate for models reduced via PEA or QSSA originates from the usage of algebraic constraints in these reduction frameworks. (3) The entropy production rate of QSSA-reduced models structurally retains contributions from all $M$ reactions of the CRN, whereas the entropy production rate derived by PEA retains only $(M-W)$ reactions. This distinction underscores a fundamental divergence between the two approximation methods: PEA eliminates fast reaction, while QSSA reduces dimensionality by constraining fast variables to algebraic equations. (4) The monotonically decreasing intrinsic Gibbs free energy $\mathcal{G}(t)$ is well preserved by QSSA but not by PEA. 

The dynamical and thermodynamic features of the original model and the reduced ones by either PEA or QSSA for open CRNs discussed above are summarized in Table \ref{table1}. These features are further illustrated through the analyses of MM reactions and phosphorylation-dephosphorylation cycle in the following sections.
\begin{table}[h]
    \centering
    \includegraphics[width=0.75\linewidth]{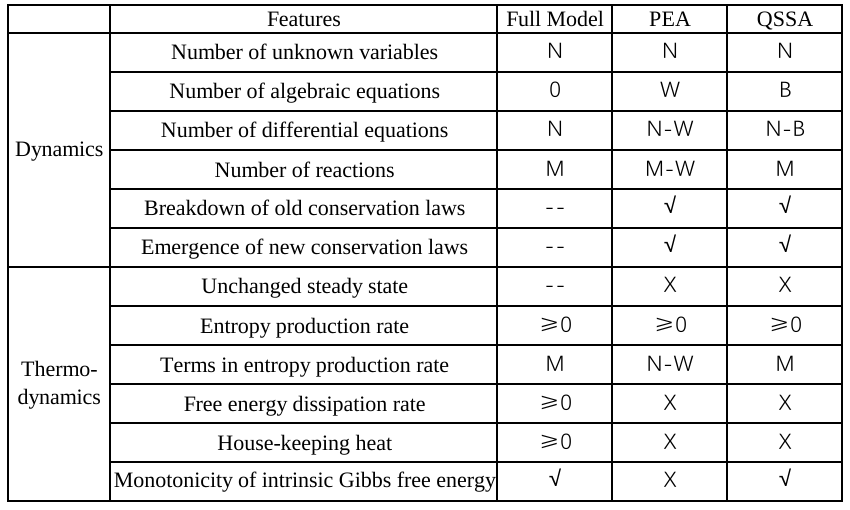}
    \caption{\textbf{Comparison of key features of the full model vs. PEA- and QSSA-reduced models for open CRNs.} \emph{Dynamical features of PEA models}: The reduced variables remain the original concentrations $\{c_k\}_{k=1}^N$, which evolve under a hybrid dynamics comprising $W$ algebraic constraints and $(N-W)$ ODEs (Eq. \eqref{PEA1and2}). The PEA framework eliminates $W$ fast reactions while retaining $(M-W)$ slow reactions. Original conservation laws may be broken as $W$ ODEs are replaced by algebraic relations, though new conservation laws can emerge too. \emph{Thermodynamic features of PEA models}: The steady state of the reduced system differs from that of the full model. The entropy production rate -- arising from $(M-W)$ slow reactions -- retains non-negativity, whereas the free energy dissipation rate loses its non-negative property The monotonically decreasing of the intrinsic Gibbs free energy is not preserved by PEA. The illustration of QSSA models follows analogously, with $B$ denoting the dimension of fast species $\bm{c}^F$.}
    \label{table1}
\end{table}

\section{Application to open Michaelis-Menten reactions}
\label{open Michaelis-Menten reactions}
The Michaelis-Menten (MM) reactions are based on the enzyme-substrate binding mechanism and have been proposed to explain the dramatic catalytic effects of enzymes on chemical reaction rates \cite{johnson2011original}. 
\begin{figure}[ht]
    \centering
    \includegraphics[width=0.75\linewidth]{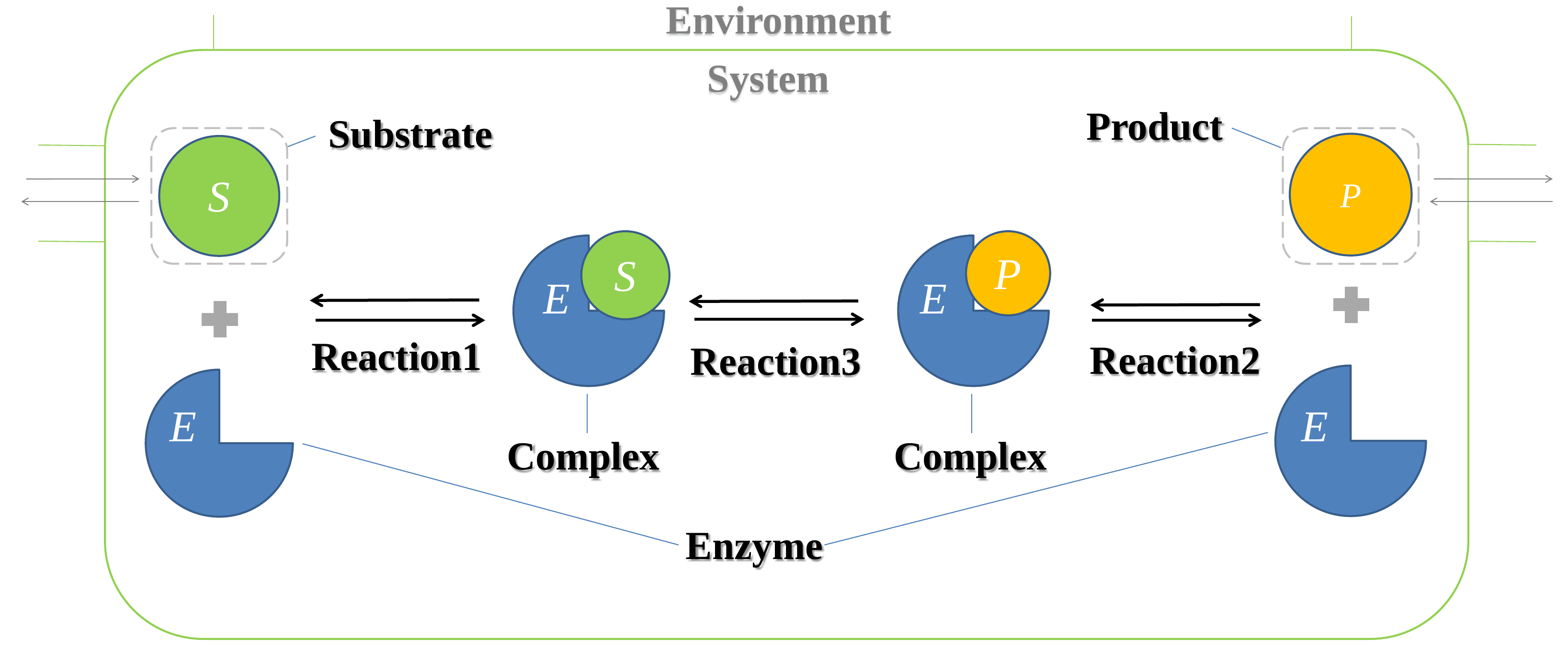}
    \caption{\textbf{Schematic representation of the open MM reactions.} The species $S$ and $P$ are chemostatted, which have continuous exchange with the environment.}
    \label{Fig_MM}
\end{figure}
In this section, we will apply the general formulation of PEA and QSSA thermodynamics to the MM reactions occurring in open systems. 

\subsection{The Full model for MM Reactions}
Here we consider an open system of MM reactions in which the substrate $S$ and product $P$ can exchange with the environment, as illustrated in Fig. \ref{Fig_MM}. The concentrations of $S$ and $P$ are assumed to be constants for simplicity.  
Given the fact that the substrate $S$ is catalyzed by enzyme $E$ to get product $P$, the MM reactions assume the existence of two intermediate complexes, $ES$ and $EP$: 
\begin{equation}
\label{MM}
\underbrace{ S{\text{ + }}E \xrightleftharpoons[{\kappa_1^-}]{{\kappa_1^+}}
}_{\text{Reaction}~1}  
\underbrace{ ES \xrightleftharpoons[{\kappa_3^-}]{{\kappa_3^+}} EP }_{\text{Reaction}~3}  
\underbrace{ \xrightleftharpoons[{\kappa_2^-}]{{\kappa_2^+}} P{\text{ + }}E }_{\text{Reaction}~2},
\end{equation}
where three reversible reactions are considered for the convenience of thermodynamic analysis, the parameters $\kappa_1^+>0$ and $\kappa_1^->0$ (resp., $\kappa_2^+, \kappa_2^-, \kappa_3^+, \kappa_3^->0$) are rate constants of the forward and backward reactions for substrate binding (resp., complex conversion, product conversion) respectively. Notice that $\kappa_3^{\pm}$ are usually assumed to be small compared to $\kappa_1^{\pm}$ and $\kappa_2^{\pm}$ , and when $\kappa_3^{\pm} \rightarrow 0$, the above mechanism degenerates into the classical two-step MM reactions.

The stoichiometric matrix $\bm \nu$ for the MM reactions \eqref{MM} is 
\begin{equation}\label{stoi}
\bordermatrix{
Reaction &S+E\rightleftharpoons ES & EP \rightleftharpoons P+E& ES\rightleftharpoons EP \cr 
        E&  -1 &   1&0 \cr 
        ES&  1 &   0& -1\cr 
        EP&  0 &  -1& 1 \cr
        S&   -1&   0&0 \cr 
        P&   0 &   1&0 
}=\bm \nu.
\end{equation}
The stoichiometric matrix $\bm \nu$ can be decomposed \cite{Rao2016Nonequilibrium} into the product of the respective composition matrix $\bm \Gamma$ and incidence matrix $\bm \Psi$ as $\bm\nu = \bm\Gamma \bm\Psi$, which are 
\begin{equation}
\label{inci_mat_MM}
  \bordermatrix{
Complex  &S+E  &  ES& EP & P+E\cr
        E&  1 &   0& 0  &1\cr 
        ES&   0 &   1& 0  &0\cr 
        EP&  0 &   0& 1  &0\cr
        S&   1 &   0& 0  &0\cr  
        P&   0 &   0& 0  &1
   }=\bm\Gamma,  
\quad
  \bordermatrix{
Reaction &S+E\rightleftharpoons ES & EP \rightleftharpoons P+E& ES\rightleftharpoons EP \cr
        S+E&  -1&   0& 0\cr 
        ES&   1 &   0&-1 \cr  
        EP&   0 &  -1& 1 \cr
        P+E&  0 &   1& 0
  }=\bm\Psi.          
\end{equation}

Denote $\bm c=([E], [ES], [EP], [S], [P])^{\dag}$. 
According to the mass-action law for the three reversible elementary reactions in \eqref{MM}, we have the net reaction rates $R_1(\bm c)=\kappa_1^+[S][E] - \kappa_1^-  [ES]$, $R_2(\bm c)=\kappa_2^+[EP] - \kappa_2^-[P][E]$ and $R_3(\bm c)=\kappa_3^+[ES] - \kappa_3^-[EP]$. Therefore, the governing equations for $\bm c$ are 
\begin{subequations}
\label{mass-action-MM}
\begin{align}
 \frac{{d[E]}}{{dt}}&=  - R_1(\bm c) + R_2(\bm c), \\ 
 \frac{{d[ES]}}{{dt}}&=  R_1(\bm c) - R_3(\bm c), \\ 
 \frac{{d[EP]}}{{dt}}&= -R_2(\bm c) + R_3(\bm c), \\
  \frac{{d[S]}}{{dt}}&=  - R_1(\bm c) + I^S(t), \\ 
 \frac{{d[P]}}{{dt}}&=  R_2(\bm c) + I^P(t), 
 \end{align}
\end{subequations}
where the initial concentrations of the enzyme $E$, complex $ES$, complex $EP$, substrate and product are assumed to be $[E]_0, [ES]_0, [EP]_0, [S]_0, [P]_0,$ respectively. 

When considering the corresponding closed MM reactions, we derive an independent set of conservation laws as $\bm l_1=(1,1,1,0,0)^{\dag}$ and $\bm l_2=(0,1,1,1,1)^{\dag}$. The conserved quantities $L\equiv {\bm l} \cdot \bm c$ are obtained as 
\begin{subequations}
    \begin{align}
\label{con1}
& \text{(Conservation law 1)}~ [E] + [ES] + [EP] = [E]_{tot}, \\
\label{con2}
& \text{(Conservation law 2)}~ [ES] + [EP]+ [S] + [P]= [S]_{tot}.
    \end{align}
\end{subequations}
Here $[E]_{tot} \equiv [E]_0 + [ES]_0 + [EP]_0$ is the total concentration of enzyme, Eq. \eqref{con1} is known as the conservation law of enzyme, and $[S]_{tot} \equiv [ES]_0 + [EP]_0+ [S]_0 + [P]_0$, Eq. \eqref{con2} states that the sum of $S$ and $P$ remains constant for the closed MM reactions. When chemostatting the species of $S$ and $P$, $\bm l_1$ is the unbroken conservation law while $\bm l_2$ is broken. 
Meanwhile, it is direct to verify that there is no internal cycle for the case in Eq. \eqref{stoi}.

We proceed to derive the complex-balanced steady state by two different approaches. First, based on the approach of regrouped complexes, all 4 complexes are regrouped into 3 sets, $\mathcal{G}_1=\{S+E, P+E\}$, $\mathcal{G}_2=\{ES\}$, $\mathcal{G}_3=\{EP\}$. 
By adding the fourth row of the matrix $\Psi$ in Eq. \eqref{inci_mat_MM} to the first row and keeping the remaining elements unchanged, we obtain the regrouped incidence matrix as 
\begin{equation}
  \bordermatrix{
Reaction &S+E\rightleftharpoons ES & EP \rightleftharpoons P+E& ES\rightleftharpoons EP \cr
        \mathcal{G}_1&  -1&   1& 0\cr 
        \mathcal{G}_2&   1 &   0&-1 \cr  
        \mathcal{G}_3&   0 &  -1& 1 
  }={\bm\Psi}^{\mathcal{G}}.         
\end{equation}
For the MM reactions in \eqref{mass-action-MM}, denote $\bm{c}^X(t) \equiv (c_1 ,c_2 ,c_3 )^{\dag} = ([E],[ES],[EP])^{\dag}$, and its steady-state, $\bm{c}_{X}^{cb} \equiv (c^{cb}_1, c^{cb}_2, c^{cb}_3)^{\dag}$. 
Therefore, the complex-balanced steady state satisfies that ${\bm\Psi}^{\mathcal{G}} \bm{R} ( \bm{c}_{X}^{cb} )=0$. In terms of the components, we have $-R_1(\bm{c}_{X}^{cb}) + R_2(\bm{c}_{X}^{cb}) = R_1(\bm{c}_{X}^{cb}) - R_3(\bm{c}_{X}^{cb}) = -R_2(\bm{c}_{X}^{cb}) + R_3(\bm{c}_{X}^{cb}) =0$, or equivalently, 
\begin{subequations}
\label{cb-MM}
\begin{align}
&- (\kappa_1 ^ +  [S][E]^{cb} - \kappa _1 ^ -  [ES]^{cb}) + (\kappa_2^+  [EP]^{cb} - \kappa_2^-  [P][E]^{cb})=0, \\ 
&  (\kappa_1 ^ +  [S][E]^{cb} - \kappa _1 ^ -  [ES]^{cb}) - (\kappa_3^+  [ES]^{cb} - \kappa_3^-  [EP]^{cb})=0, \\ 
& -(\kappa_2^+  [EP]^{cb} - \kappa_2^-  [P][E]^{cb}) + (\kappa_3^+  [ES]^{cb} - \kappa_3^-  [EP]^{cb})=0.
 \end{align}
\end{subequations}
There exists a unique positive complex-balanced steady state $\bm{c}^s >0$ that satisfies Eq. \eqref{cb-MM}: 
\begin{subequations}
\label{ness-MM}
\begin{align}
 [E]^{cb}&=  \frac{E_{tot}}{\mathcal{D}}(\kappa_1^- \kappa_2^+  + \kappa_1^- \kappa_3^-  + \kappa_2^+ \kappa_3^+ ), \\ 
 [ES]^{cb}&=  \frac{E_{tot}}{\mathcal{D}}(\kappa_1^+\kappa_2^+[S]  + \kappa_1^+\kappa_3^-[S]  + \kappa_2^-\kappa_3^-[P] ), \\ 
[EP]^{cb}&= \frac{E_{tot}}{\mathcal{D}}(\kappa_1^- \kappa_2^-[P]  + \kappa_1^+\kappa_3^+[S]  + \kappa_2^-\kappa_3^+[P] ),
 \end{align}
\end{subequations}
where the denominator $\mathcal{D}=\kappa_1^- \kappa_2^+  + \kappa_1^- \kappa_3^-  + \kappa_2^+ \kappa_3^+ + \kappa_1^+\kappa_2^+[S] + \kappa_1^+\kappa_3^-[S]  + \kappa_2^- \kappa_3^-[P] + \kappa_1^- \kappa_2^-[P]  + \kappa_1^+\kappa_3^+[S]  + \kappa_2^- \kappa_3^+[P]$. 

On the other hand, by absorbing the effects of chemostats into pseudo reaction rates, and thus treating open CRNs as effective closed ones \cite{ge2013dissipation, qian2021stochastic}, we arrive at an effective representation of the open MM reactions  as
\begin{equation}
\label{pseudo MM}
\schemestart
    \chemname{$ES$}{} 
    \arrow{<=>[$\kappa_3^+$][$\kappa_{3}^-$]}[,1.4]
    \chemname{$EP$}{} 
    \arrow{<=>[$\kappa_2^-[P]$][$\kappa_{2}^+$]}[130,1]
    \chemname{$E$}{} 
    \arrow{<=>[$\kappa_1^-$][$\kappa_{1}^+[S]$]}[-130,1]
\schemestop
\end{equation}
This effective representation provides another method to derive the complex-balanced steady state. For the open MM reactions in \eqref{pseudo MM}, an equivalent steady state can be deduced by simply letting the current of each cycle being zero. That is, $\kappa_1 ^ +  [S][E]^{cb} - \kappa _1 ^ -  [ES]^{cb} = \kappa_2^+  [EP]^{cb} - \kappa_2^-  [P][E]^{cb}=\kappa_3^+  [ES]^{cb} - \kappa_3^-  [EP]^{cb}$, which will lead to the same results as in Eq. \eqref{ness-MM}.

According to the general theory of nonequilibrium thermodynamics for CRNs presented in Sect. \ref{noneq ther}, we can derive the thermodynamic quantities for the full MM reactions, including the enthalpy, entropy, intrinsic Gibbs free energy, and relative energy function, as well as their time change rates and further decompositions. See \textcolor{blue}{SI}
for details. 

\subsection{MM Reactions Reduced by PEA}
\label{MM PEA therm}
Here we assume the association and disassociation of the substrate and enzyme (Reaction 3) proceed in a time scale much shorter than Reactions 1 and 2, which means Reaction 3, $ES \xrightleftharpoons[{\kappa_3^-}]{{\kappa_3^+}} EP$, is considered to be fast. By  explicitly writing out the order $\epsilon$ of MM reactions, we have 
\begin{subequations}
\label{PEA-MM}
\begin{align}
 \frac{{d[E]}}{{dt}}&=  -R_1(\bm{c}) + R_2(\bm{c}), \\ 
 \frac{{d[ES]}}{{dt}}&=  R_1(\bm{c}) - \frac{1}{\epsilon} \widehat{R_3}(\bm{c}), \\ 
 \frac{{d[EP]}}{{dt}}&= -R_2(\bm{c}) + \frac{1}{\epsilon} \widehat{R_3}(\bm{c}), 
 \end{align}
\end{subequations}
where $\widehat{R_3}(\bm{c}) \equiv \epsilon R_3(\bm{c})$ are of the same order as ${R_1}(\bm{c})$. 

Taking the limit $\epsilon \rightarrow 0$, we have the algebraic equation, called the PEA relation $R_3(\bm{c}) = 0$, or equivalently, 
\begin{equation}
\label{R1pm}
\kappa _3^+[ES] = \kappa _3^-[EP].
\end{equation}
In this stage, we have 3 reduced ODEs ${{d[E]}}/{{dt}}=  -R_1(\bm{c}) + R_2(\bm{c})$, ${{d[ES]}}/{{dt}}=R_1(\bm{c})$, and ${{d[EP]}}/{{dt}}=-R_2(\bm{c})$ for 3 variables with the corresponding initial values $([E]_0, [ES]_0, [EP]_0)^{\dag}$ and an extra PEA relation $R_3(\bm{c}) =0$. These 4 equations constitute an over-determined system for 3 unknowns. 

As to this case, there is only one fast reaction, $W=1$, and the concentration $[EP]$ is expressed by that of $[ES]$ as $[EP]=k_3^+ [ES]/k_3^-$, which indicates that $V=1$. Following the general theory of PEA established above, we neglect the ODE $d[EP]/dt=0$ and obtain the PEA-reduced dynamics of the MM reactions in terms of the state variables $([E], [ES])^{\dag}$ as
\begin{subequations}
\label{PEA-MM1}
\begin{align}
 \frac{{d[E]}}{{dt}}&=  -R_1(\overline{\bm{c}}) + R_2(\overline{\bm{c}}), \\ 
 \frac{{d[ES]}}{{dt}}&=  R_1(\overline{\bm{c}}),
\end{align}
\end{subequations}
here the initial values are $([E]_0, [ES]_0)^{\dag}$, and $R_1(\overline{\bm{c}})=\kappa _1 ^ +[S][E] - \kappa_1^-  [ES]$, $R_2(\overline{\bm{c}})=\frac{\kappa_2^+  k_3^+}{k_3^-} [ES] - \kappa_2^- [P][E]$. The concentration $[EP]=k_3^+ [ES]/k_3^-$ is obtained by substitution of the solutions to the above ODEs. 

By setting the right-hand side of the PEA-reduced MM model in \eqref{PEA-MM1} to $0$, we obtain a constrain on the rate constants and the chemostatted concentrations as 
\begin{equation}
\frac{\kappa_1^+ \kappa_2^+ \kappa_3^+ [S]}{\kappa_1^- \kappa_2^- \kappa_3^-[P]} =1,
\end{equation}
which is known as the detailed balance condition for the original MM reactions. {This condition guarantees the existence of the steady state of the PEA-reduced MM model \eqref{PEA-MM1}.} However, due to the inclusion of the PEA relation, the PEA-reduced MM model \eqref{PEA-MM1} is no longer subject to the law of mass action, and the conservation law 1 is also broken as $[E] + [ES] + [EP] \neq [E]_{tot}$. 
The thermodynamic quantities for the PEA-reduced MM model are readily obtained in \textcolor{blue}{SI}.

\subsection{MM Reactions Reduced by QSSA}
\label{QSSA_thermodynamics_MM}
Different from PEA, QSSA assumes that the synthesis and decomposition rates of the complex $ES$ equal, or equivalently, the second ODE in \eqref{mass-action-MM} is reduced to an algebraic equation, 
\begin{equation}
\label{QSSA_MM}
R_1(\widetilde{\bm c}) = R_3(\widetilde{\bm c}).
\end{equation} 
In this stage, we have 2 reduced ODEs 
\begin{subequations}
\label{mass-action-MM4}
\begin{align}
 \frac{{d[E]}}{{dt}}&=  - R_1(\widetilde{\bm c}) + R_2(\widetilde{\bm c}), \\
 \frac{{d[EP]}}{{dt}}&= -R_2(\widetilde{\bm c}) + R_3(\widetilde{\bm c}), 
 \end{align}
\end{subequations}
with the initial concentrations $([E]_0, [EP]_0)^{\dag}$ and the QSSA relation in Eq. \eqref{QSSA}. 
Notice that by applying QSSA to MM reactions, all old conservation laws have been broken. Instead, a new conservation law emerges, $[E]+[EP]=\widetilde{E_{tot}}$ with $\widetilde{E_{tot}} \equiv [E]_0+[EP]_0$, due to the QSSA relation in \eqref{QSSA_MM}. 

By the QSSA relation in \eqref{QSSA_MM}, we obtain the expression of the complex $ES$ represented via the enzyme $E$ and the other complex $EP$, 
\begin{equation}
\label{CE}
 [ES] = \frac{\kappa_1^+ [S][E] + \kappa_3^- [EP]}{\kappa_1^- + \kappa_3^+}.
\end{equation}
Upon substituting the above algebraic formula of $[ES]$ into the ODEs in \eqref{mass-action-MM4}, we obtain the closed form of QSSA-reduced dynamics for the MM reactions as 
\begin{subequations}
\label{QSSA-MM1}
\begin{align}
 \frac{{d[E]}}{{dt}}&= -(\kappa_2^-[P] +\frac{\kappa_1^+ \kappa_3^+}{\kappa_1^+ + \kappa_3^+}[S]) [E] + (\kappa_2^+ + \frac{\kappa_1^- \kappa_3^-}{\kappa_1^- + \kappa_3^+} )[EP], \\
 \frac{{d[EP]}}{{dt}}&= (\kappa_2^-[P] + \frac{\kappa_1^+ \kappa_3^+}{\kappa_1^- + \kappa_3^+}[S]) [E] - (\kappa_2^+ + \frac{\kappa_1^- \kappa_3^-}{\kappa_1^- + \kappa_3^+})[EP], 
 \end{align}
\end{subequations}
where the initial condition is $([E]_0, [EP]_0)^{\dag}$. 
The Eqs. \eqref{CE}-\eqref{QSSA-MM1} together are called the QSSA-MM model, whose steady state is given by 
\begin{subequations}
\label{ness-QSSA-MM1}
\begin{align}
  \widetilde{[E]}^{ss}&=  \frac{\widetilde{E_{tot}}}{\widetilde{\mathcal{D}}}(\kappa_1^- \kappa_2^+  + \kappa_1^- \kappa_3^-  + \kappa_2^+ \kappa_3^+ ), \\ 
 \widetilde{[ES]}^{ss}&= \frac{\widetilde{E_{tot}}}{\widetilde{\mathcal{D}}}(\kappa_1^+ \kappa_2^+[S]  + \kappa_1^+ \kappa_3^-[S]  + \kappa_2^- \kappa_3^-[P] ),\\
 \widetilde{[EP]}^{ss}&= \frac{\widetilde{E_{tot}}}{\widetilde{\mathcal{D}}}(\kappa_1^- \kappa_2^-[P]  + \kappa_1^+ \kappa_3^+[S]  + \kappa_2^- \kappa_3^+[P] ),
 \end{align}
\end{subequations}
where the denominator $\widetilde{\mathcal{D}}=\kappa_1^- \kappa_2^+  + \kappa_1^- \kappa_3^-  + \kappa_2^+ \kappa_3^+ + \kappa_1^- \kappa_2^-[P]  + \kappa_1^+ \kappa_3^+[S] + \kappa_2^- \kappa_3^+[P]$. 
This steady state is usually different from that of the full model, in Eq. \eqref{ness-MM}.

The fast and slow species are $\widetilde{\bm{c}}^F=[ES]$ and $\widetilde{\bm{c}}^S=([E],[EP],[S],[P])^{\dag}$ separately. The stoichiometric matrix $\bm \nu$ for the MM reactions is decomposed into 
$\bm{\nu}^F=\begin{pmatrix}
 1 & 0 &-1
\end{pmatrix}$ 
and 
$$\bm{\nu}^S=
\begin{pmatrix}
  -1&  1&0 \\
 0 & -1 &1 \\
 -1 &0  &0 \\
 0 & 1 &0
\end{pmatrix}.$$ 
We define the right null eigenvectors of $\bm{\nu}^F$ as $\bm{\phi}_{\gamma}$, which satisfies $\bm{\nu}^F\bm{\phi}_{\gamma}=\bm{0}$. Then, for the MM reactions, we derive two independent right null eigenvectors as $\bm{\phi}_{1}$=${(1,0,1)^{\dag}}$, $\bm{\phi}_{2}$=${(1,1,1)^{\dag}}$, both of which are pseudo-emergent cycles since $\bm{\nu}^F\bm{\phi}_{\gamma} \neq \bm{0}$. Recall that the tilde $\widetilde{\mathcal{C}}$ is utilized to denote the QSSA-reduced results. Therefore, the effective stoichoimetric matrix becomes 
\begin{equation}
\bm{\widetilde \nu}^S=\bm{ \nu}^S(\bm{\phi}_{1},\bm{\phi}_{2})=\begin{pmatrix}
  -1&0 \\
  1&0 \\
  -1&-1 \\
  0&1
\end{pmatrix}.
\end{equation}
In this way, we can obtain the effective dynamics:
\begin{equation}
\frac{d}{dt}\begin{pmatrix}
 [E]\\
 [EP]\\
 [S]\\
[P]
\end{pmatrix} =\bm{\widetilde \nu}^S\begin{pmatrix}
 \psi _{1} \\
\psi _{2} 
\end{pmatrix}+\begin{pmatrix}
 0\\
 0\\
 I^{S}(t) \\
I^{P}(t)
\end{pmatrix}, 
\end{equation}
where the coefficients are 
\begin{subequations}
\begin{align}
\psi_{1}(\widetilde{\bm c})&
=\frac{k_{1}^{+}k_{3}^{+}[S][E] - k_{1}^{-}k_{3}^{-}[EP] }{k_{1}^{-}+k_{3}^{+}} - (k_{2}^{+}[EP]-k_{2}^{-}[P] [E]), \\
\psi_{2}(\widetilde{\bm c})&=k_{2}^{+}[EP]-k_{2}^{-}[P] [E]. 
\end{align}
\end{subequations}
Correspondingly, the effective MM reactions are represented as  
\begin{equation}
\label{r_effective MM}
S+E \xrightleftharpoons[{}]{{\psi _{1}}} EP, 
\quad 
S \xrightleftharpoons[{}]{{\psi _{2}}} P.
\end{equation}

Recall that the unbroken conservation law and broken conservation law for the original dynamics are $\bm{u}=\bm {\ell}_1=(1,1,1,0,0)^{\dag}$ and $\bm{b}=\bm {\ell}_2=(0,1,1,1,1)^{\dag}$. 
As to the QSSA-reduced effective dynamics, the unbroken conservation law is $\bm{\tilde{u}}=\mathbb{P}\bm{u}=(1,1,0,0)^{\dag}$, while the broken conservation law is $\bm{\tilde{b}}=\mathbb{P} \bm{b}=(0,1,1,1)^{\dag}$, where $\mathbb{P}$ stands for the projection operator. 
As mentioned earlier, the chemostatted species are $\bm{Y}=(S,P)^{\dag}$. Since chemostatting a species does not always break a conservation law, we thus distinguish the set of controlled species $\bm{Y}_p$ breaking the conservation laws from the rest $\bm{Y}_f=\bm{Y}\setminus \bm{Y}_p$. Therefore, the number of $\bm{Y}_p$ for the MM reactions equals to that for the broken conservation laws, $|\lambda _{b}|=1$. Here we choose $\bm{Y}_p={P}$ and $\bm{Y}_f={S}$.

\emph{Remark:} 
When chemostatting the species $S$ and $P$, an independent emergent cycle arises, that is, $\bm{\nu}^X \bm{c}_{\varepsilon }=\bm{0}$, where $\bm{\nu}^X$ is the stoichiometric matrix for internal species. For the MM reactions,
$$\bm{\nu}^{X} =\begin{pmatrix}
 1 & 0 &-1 \\
 -1 &1  &0 \\
  0& -1 &1
\end{pmatrix}.$$
It is easy to obtain $\bm{c_{\varepsilon }}=(1,1,1)^{\dag}$. Thus, the number of independent emergent cycle $\bm{|\varepsilon |}=1$. Meanwhile, the number of broken conservation laws is $|\lambda _{b}|=1$, and the number of chemostatted species is $|S _{Y}|=2$. This verifies the topological properties $|S _{Y}|=\bm{|\varepsilon |}+|\lambda _{b}|$. See Ref.\cite{Rao2016Nonequilibrium} for details. 

The thermodynamic quantities for the QSSA-reduced MM model are presented in \textcolor{blue}{SI}.
We also noticed that there is a similar work \cite{Avanzini2020}  discussing the thermodynamics for non-elementary reactions, especially for QSSA-reduced CRNs. We applied this thermodynamic framework to the open MM reactions and derived the corresponding quantities for the simplified model (See 
Appendix \ref{Derivation of the thermodynamic quantities in the framework}
for details). By comparing these thermodynamic quantities with ours, we conclude that: (1) Our work establishes a thermodynamic framework for reduced models of open CRNs that explicitly incorporates both the first and second laws of thermodynamics along with a comprehensive set of thermodynamic quantities. In contrast, Ref. \cite{Avanzini2020} primarily addresses topological considerations, as well as the entropy production rate and Gibbs free energy. (2) The equivalence between the entropy production rates expressed in Eqs. \eqref{thermo_QSSA} and \eqref{thermo_Avan} emerges naturally from the local detailed balance condition. (3) Notably, the intrinsic Gibbs free energy presented in our main text differs fundamentally from the semigrand Gibbs free energy approach developed in \cite{Avanzini2020}, reflecting distinct thermodynamic perspectives.

\subsection{Numerical Illustration}
\begin{figure}[h]
    \centering
    \includegraphics[width=0.75\linewidth]{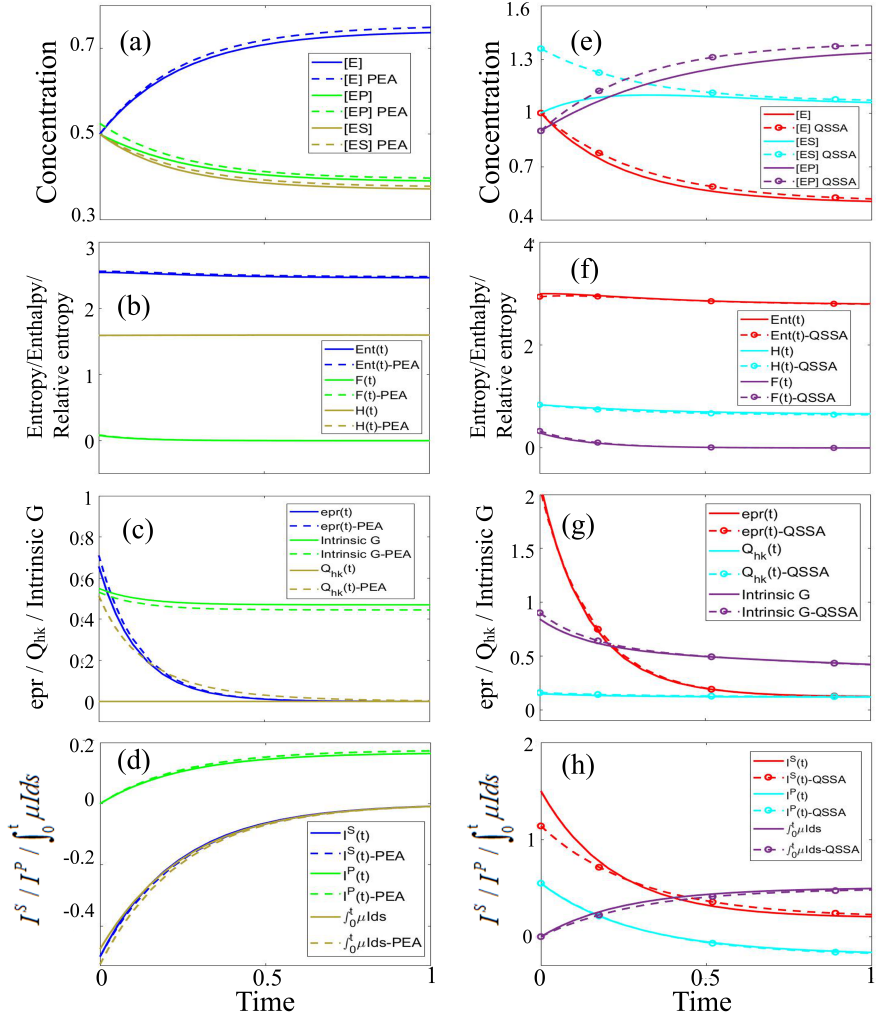}
    \caption{\textbf{Illustration on the dynamics and thermodynamics of the full MM reactions (solid lines) and reduced models by PEA  and QSSA (dashed lines).} From the left four panels (a-d) to the right panels (e-h), the results for PEA and QSSA models are shown respectively.   
    (a, e) The concentrations of enzyme $[E]$, complex $[ES]$, and complex $[EP]$ for the complete and reduced models are compared and plotted. So are the entropy $Ent(t)$, enthalpy $H(t)$, free energy $F(t)$ in (b, f); the entropy production rate $epr(t)$, house-keeping heat $Q_{hk}(t)$, and intrinsic Gibbs free energy $\mathcal{G}(t)$ in (c, g). 
    The external currents  $I^S(t), I^P(t)$, and the integral part of intrinsic Gibbs free energy $\int_0^t \mu I ds$ are plotted in (d, h). 
    In all plots, the initial values and rate constants are taken as $([S]_0, [ES]_0, [EP]_0) = (0.5, 0.5, 0.5)$, $(\kappa_1^+, \kappa_1^-, \kappa_2^+, \kappa_2^-, \kappa_3^+, \kappa_3^-)=(1, 2, 2, 1.05, 3, 3/1.05)$ for (a-d); and $([S]_0, [ES]_0, [EP]_0) = (1, 1, 0.9)$, $(\kappa_1^+, \kappa_1^-, \kappa_2^+, \kappa_2^-, \kappa_3^+, \kappa_3^-)=(2.5, 1, 0.5, 1, 1.5, 1)$ for (e-h) correspondingly. We set the chemostatted species $[S]\equiv[P]\equiv 1$, the standard entropy of formation $\bm s^{\circ}=\bm 0$, the standard enthalpy of formation $\bm h^{\circ}$ based on the local detailed balance condition, and the  constants $\mathcal{R}=\mathcal{T}=1$.}
    \label{Fig_MM_thermo}
\end{figure}

To provide an intuitive understanding of our thermodynamics for the reduced models by PEA and QSSA, we perform numerical calculations of the MM reactions as an illustration. As shown by the trajectories of the enzyme $[E](t)$, complexes $[ES](t)$ and $[EP](t)$ in Fig. \ref{Fig_MM_thermo}(a,e), the reduced models by either PEA or QSSA offer quite good approximations on original solutions to the full model, except for the initial layer of $[ES]$ for the QSSA model. 
As to the thermodynamic behaviors of the reduced models which we are more concerned about, we make an exploration based on three thermodynamic state functions: entropy $Ent(t)$, enthalpy $H(t)$ and relative entropy $F(t)$ in Fig. \ref{Fig_MM_thermo}(b,f). They change more gently over time than the concentrations, thanks to the characterization of the CRN as a whole by the thermodynamic quantities. 
In Fig. \ref{Fig_MM_thermo}(c), the housekeeping heat of the full model equals to zero, $Q_{hk}(t)\equiv 0$, while $\overline{Q_{hk}}(t) \neq 0$ for the PEA model due to the breakdown of the detail-balanced condition. When moving to the case in Fig. \ref{Fig_MM_thermo}(g), both the full model and the QSSA-reduced model exhibit positive housekeeping heat, which is an important feature of open CRNs different from closed ones. Moreover, the intrinsic Gibb's free energy $\mathcal{G}(t)$, and the terms $I(t), \int_0^t \mu(s) I(s) ds$ in Fig. \ref{Fig_MM_thermo}(d,h) are also characteristics of open CRNs. 


\section{Application to protein phosphorylation-dephosphorylation cycle}
\begin{figure}[ht]
    \centering
    \includegraphics[width=0.75\linewidth]{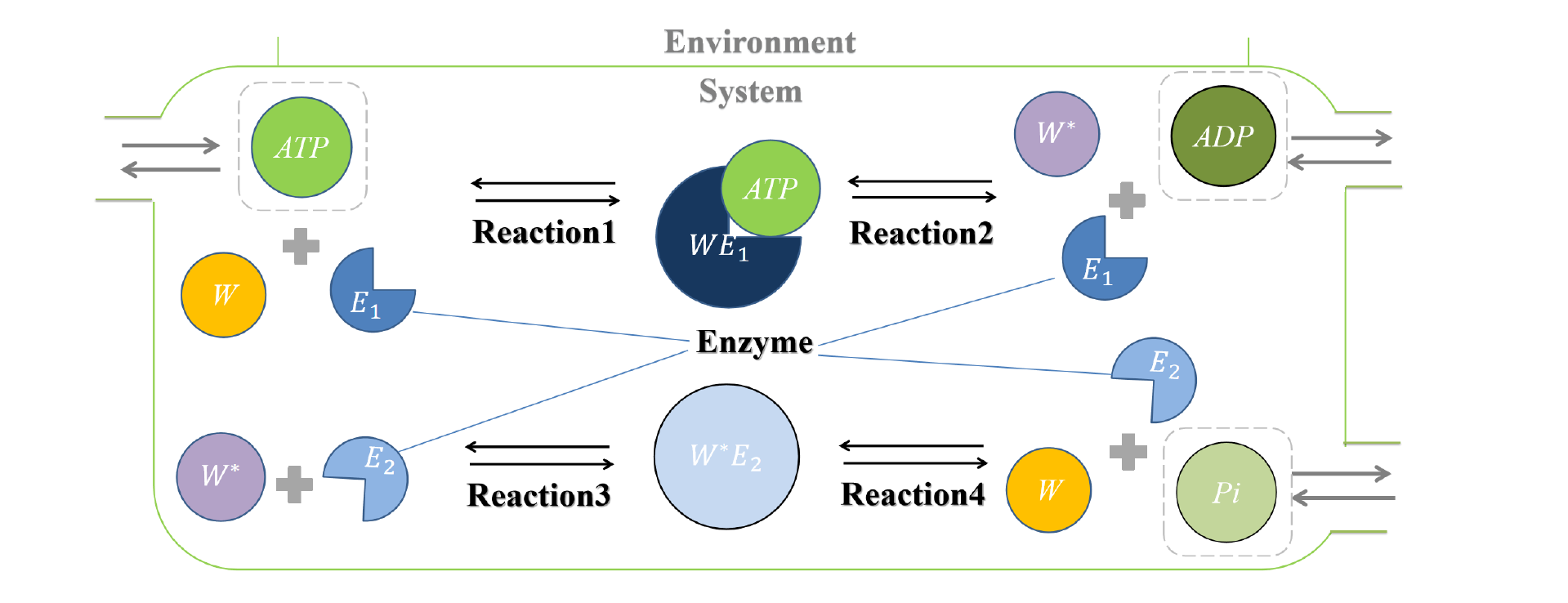}
    \caption{\textbf{Schematic representation of the open PdPC reactions.} The species $\mathrm{ATP}$, $\mathrm{ADP}$ and $\mathrm{Pi}$ are chemostatted, which have continuous exchange with the environment.}
    \label{Fig_PdPC}
\end{figure}

Biological signal transduction processes are increasingly understood in modular and quantitative terms. A key module in cellular circuitry that has been extensively studied is the protein phosphorylation-dephosphorylation cycle (PdPC, shown in Fig. \ref{Fig_PdPC}), which demonstrates significant amplification of sensitivity in response to appropriate stimuli, achieved through the activation of a kinase or inhibition of a phosphatase \cite{Krebs1981401}.

In this section, we will apply the QSSA method to simplify a reversible model of PdPC \cite{Qian2003}. The system dynamics, thermodynamics, and sensitivity before and after simplification will be compared in detail to provide a clear understanding of the system’s behaviors and their impacts on energy consumption and reaction rates. 

\subsection{The Full model for PdPC Reactions}
\label{sec:PdPC_full}

A model for protein phosphorylation-dephosphorylation cycle modulated by two different kinds of enzymes $E_1$ and $E_2$ reads
\begin{equation}
    \label{PdPC_crn}
    W+E_{1}+\mathrm{ATP} \underset{\kappa_{1}^-}{\stackrel{\kappa_{1}^{+o}}{\rightleftharpoons}} W \cdot E_{1} \cdot \mathrm{ATP} \underset{\kappa_{2}^{-o}}{\stackrel{\kappa_{2}^+}{\rightleftharpoons}} W^{*}+E_{1}+\mathrm{ADP},
    \quad 
    W^{*}+E_{2} \underset{\kappa_{3}^-}{\stackrel{\kappa_{3}^+}{\rightleftharpoons}}  W^{*} E_{2}  \underset{\kappa_{4}^{-o}}{\stackrel{\kappa_{4}^+}{\rightleftharpoons}} W+E_{2}+\mathrm{Pi}.
\end{equation}
Here the concentrations of \textit{ATP}, \textit{ADP}, and \textit{Pi} are assumed to be constant because these substances are in dynamic equilibrium with the body rapidly, regulating their levels through various metabolic processes to maintain homeostasis. Consequently, we can combine them with the reaction rate constants as
$\kappa_1^+= \kappa_1^{+o} [\text{ATP}], 
\kappa_{2}^{-} = \kappa_{2}^{-o} [\text{ADP}], 
\kappa_{4}^{-} = \kappa_{4}^{-o} [\text{Pi}]$. 
By denoting \([WE_1]=[W \cdot E_{1} \cdot A T P]\), we obtain a simplified CRN as $W+E_{1}\underset{\kappa_{1}^-}{\stackrel{\kappa_{1}^+}{\rightleftharpoons}} W E_{1} \underset{\kappa_{2}^-}{\stackrel{\kappa_{2}^+}{\rightleftharpoons}} W^{*}+E_{1}, ~W^{*}+E_{2} \underset{\kappa_{3}^-}{\stackrel{\kappa_{3}^+}{\rightleftharpoons}} W^{*} E_{2} \underset{\kappa_{4}^-}{\stackrel{\kappa_{4}^+}{\rightleftharpoons}} W+E_{2}$. Consequently, the dynamics of PdPC in \eqref{PdPC_crn} is also governed by the mass-action equations in \eqref{massactioneq_0} with the concentration vector, the stoichiometric matrix, the external current vector, and the reaction rate vectors defined as: 
\begin{equation}
\label{PdPC_defs}
\bm c=\begin{pmatrix}
    [E_1]\\
    [WE_1]\\
    [W]\\
    [W^*]\\
    [E_2]\\
    [W^*E_2]\\
    [\mathrm{ATP}]\\
    [\mathrm{ADP}]\\
    [\mathrm{Pi}]
\end{pmatrix},
\quad
\bm \nu=\begin{pmatrix}
        -1 &1&0 & 0\cr
        1&  -1 &  0&  0\cr
        -1 &   0&0&1 \cr 
        0&1 & -1&   0\cr 
        0&0 &   -1&   1 \cr
        0&0 &   1&   -1\cr 
        -1 &  0 &   0&0 \cr
        0 &   1&0 &0\cr
        0 &0&0&   1
\end{pmatrix}, 
\quad
\bm I =\begin{pmatrix}
    0\\
    0\\
    0\\
    0\\
    0\\
    0\\
    I_{\mathrm{ATP}}\\
    I_{\mathrm{ADP}}\\
    I_{\mathrm{Pi}}
\end{pmatrix},
\quad
\bm R^+ =\begin{pmatrix}
\kappa_1^+ [W][E_1]\\
\kappa_2^+ [W E_1]\\
\kappa_3^+ [W^*][E_2]\\
\kappa_4^+ [W^*E_2]
\end{pmatrix},
\quad
\bm R^- =\begin{pmatrix}
\kappa_1^- [W E_1]\\
\kappa_2^- [W^*][E_1]\\
\kappa_3^- [W^*E_2]\\
\kappa_4^- [W][E_2]
\end{pmatrix}.
\end{equation}
Here \( I_{\mathrm{ATP}} = R_1 \), \( I_{\mathrm{ADP}} = -R_2 \), and \( I_{\mathrm{Pi}} = R_4 \) due to the fact that \([\mathrm{ATP}]\), \([\mathrm{ADP}]\), and \([\mathrm{Pi}]\) remain constant.

There exist three conservation laws for the PdPC reactions, that is 
\begin{subequations}
    \begin{align}
&[W]+\left[W^{*}\right]+\left[W E_{1}\right]+\left[W^{*} E_{2}\right]=[W]_{tot} ,\label{PdPC_cons1}\\
&\left[E_{1}\right]+\left[W E_{1}\right]=[E_1]_{tot},\label{PdPC_cons2} \\
&\left[E_{2}\right]+\left[W^{*} E_{2}\right]=[E_2]_{tot},
\label{PdPC_cons3}
\end{align}
\end{subequations}
where $[W]_{tot}$, $[E_1]_{tot}$ and $[E_2]_{tot}$ stand for the total concentrations of $W$, $E_1$ and $E_2$ separately. Based on the above dynamic equations, the thermodynamic quantities for the full PdPC reactions can be derived analogously as the MM reactions (see SI for details).

\subsection{PdPC Reactions Reduced by QSSA}
In this part, we assume the enzyme $E_1$ is maintained at a dynamical equilibrium. Thus the QSSA implies that the ODE \(d[E_1]/dt =-R_1(\bm c)+R_2(\bm c)\) is replaced by an algebraic equation \(R_1(\bm c)=R_2(\bm c)\), that is 
\begin{equation}
    \widetilde{[E_1]} = \frac{(\kappa_1^- + \kappa_2^+)\widetilde{[WE_1]}}{\kappa_1^+\widetilde{[W]} + \kappa_2^-\widetilde{[W^*]}}.
    \label{PdPC_QSSA}
\end{equation}
By making use of above formula, the governing equations for the QSSA-reduced PdPC model become
\begin{subequations}
\begin{align}
    &\frac{d\widetilde{\bm{c}}^S}{dt} = \bm{\nu}^S \bm{R}(\widetilde{\bm{c}}) + \bm{I}^S, \\
    &{\bm{\nu}}^F \bm{R}(\widetilde{\bm{c}}) = 0,
\end{align}
\label{PdPC_mae_QSSA}
\end{subequations}
where \({\bm{\nu}}^F\) refers to the first row of \(\bm{\nu}\), and \({\bm{\nu}^S}\) refers to the remaining rows. Furthermore, we have
\begin{subequations}
\begin{align}
\label{PdPC_defs_QSSA}
\widetilde{\bm{c}}^F &=(\widetilde{[E_1]}), \quad
\widetilde{\bm{c}}^S =(\widetilde{[WE_1]},\widetilde{[W]},\widetilde{[W^*]},\widetilde{[E_2]},\widetilde{[W^*E_2]},\widetilde{[\mathrm{ATP}]},\widetilde{[\mathrm{ADP}]},\widetilde{[\mathrm{Pi}]})^{\dag},\\ 
\bm{I}^F &=(0),  \quad
\bm{I}^S =(0,0,0,0,0,I_{\mathrm{ATP}},I_{\mathrm{ADP}},I_{\mathrm{Pi}})^{\dag}.
\end{align}
\end{subequations}
It is noticeable that the first and third conservation laws \eqref{PdPC_cons1} and \eqref{PdPC_cons3} are preserved in the QSSA-reduced model, while the second one \eqref{PdPC_cons2} has been broken. 

\begin{figure}[h]
    \centering
    \includegraphics[width=0.9\linewidth]{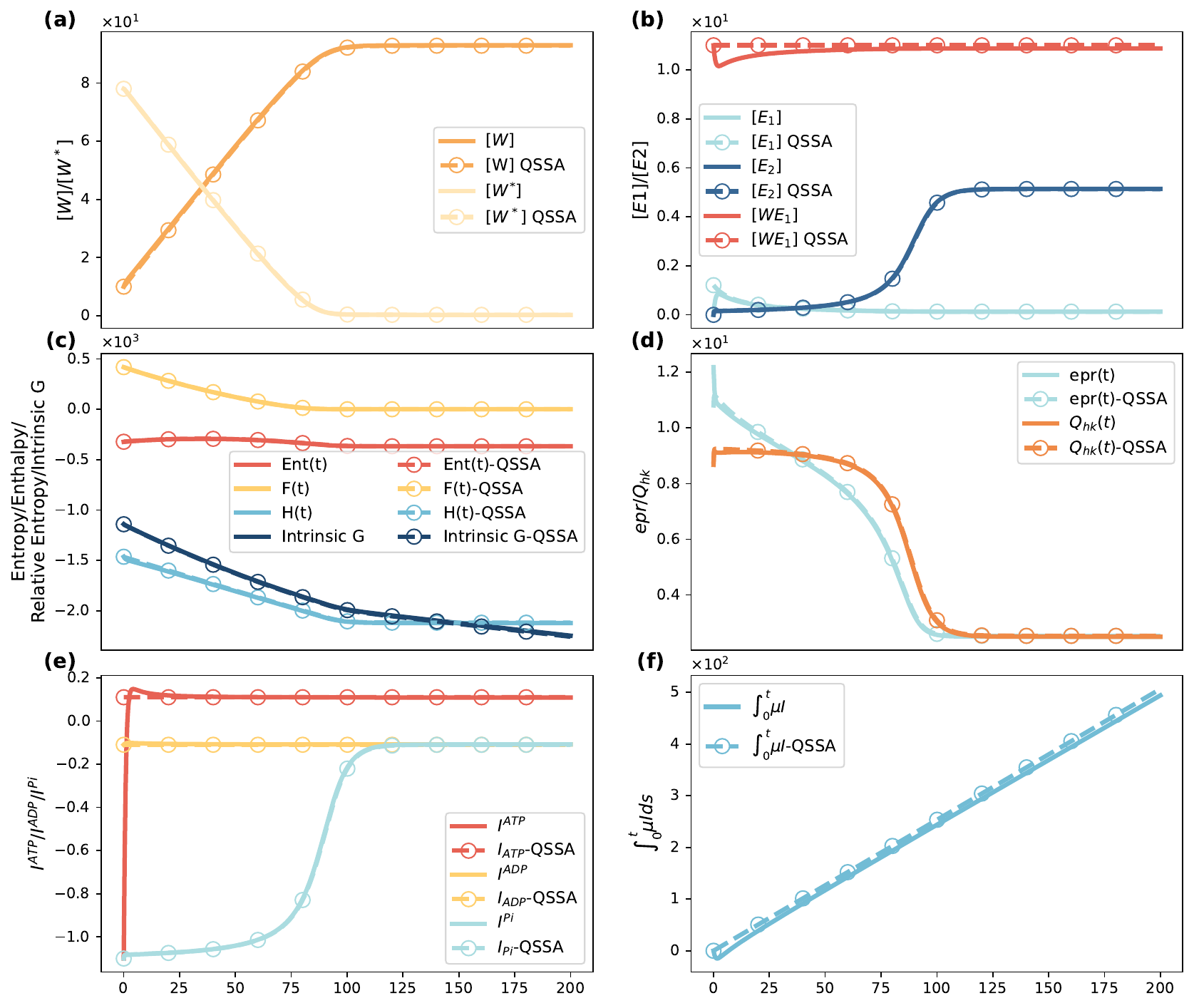}
    \caption{\textbf{Illustration on the dynamics and thermodynamics of the full PdPC reactions (solid lines) and QSSA-reduced models (dashed lines).} (a, b) The concentrations of enzyme $[E_1](t),[E_2](t)$ and protein $[W](t),[W^*](t)$ of the full and reduced models are compared and plotted. So are the entropy $Ent(t)$, enthalpy $H(t)$, free energy $F(t)$ , and intrinsic Gibbs free energy $\mathcal{G}(t)$ in (c); the entropy production rate $epr(t)$, housekeeping heat  $Q_{hk}(t)$ in (d). The complex $[W^*E_2](t)$ and enzyme $[E_2](t)$ satisfy the conservation law in \eqref{PdPC_cons3}. The external currents  $I^{\mathrm{ATP}}(t), I^{\mathrm{ADP}}(t),I^{\mathrm{Pi}}(t)$, and the integral part of intrinsic Gibbs free energy $\int_0^t \mu I ds$ are plotted in (e, f). 
    For all plots, the initial values and rate constants are taken as $([E_1]_0, [WE_1]_0, [W]_0, [W^*]_0,[E_2]_0,[W^*E_2]_0,[\mathrm{ATP}]_0,[\mathrm{ADP}]_0,[\mathrm{Pi}]_0) = (0, 11, 10, 78, 0, 11, 1, 1, 1)$, $(\kappa_1^+, \kappa_1^-, \kappa_2^+, \kappa_2^-, \kappa_3^+, \kappa_3^-,\kappa_4^+, \kappa_4^-)=(10^{-1}, 10^{-1}, 10^{-2}, 10^{-9}, 10^{-1}, 10^{-1}, 10^{-1}, 10^{-3})$. We set the standard entropy of formation $\bm s^{\circ}=\bm 0$, the standard enthalpy of formation $\bm h^{\circ}=(0,0,-\ln{\frac{\kappa_2^+\kappa_3^+\kappa_4^+}{\kappa_2^-\kappa_3^-\kappa_4^-}},-\ln{\frac{\kappa_2^+}{\kappa_2^-}},0,0,\ln{\frac{\kappa_1^+\kappa_2^+\kappa_3^+\kappa_4^+}{\kappa_1^-\kappa_2^-\kappa_3^-\kappa_4^-}},0,0)$ based on the local detailed balance condition. The constant $\mathcal{R}=\mathcal{T}=1$.}
    \label{Fig_PdPC_QSSA}
\end{figure}

All thermodynamic quantities for the QSSA-reduced PdPC model can be derived based on the general theory in Sect. \ref{QSSA thermodynamics}, which are illustrated in the following. Figs. \ref{Fig_PdPC_QSSA}(a)-(f) collectively demonstrate that the QSSA model approximates the original PdPC model quite well. 
As fast species in the QSSA model, the value of \(\widetilde{[E_1]}\) is determined by \(\widetilde{[WE_1]}\), \(\widetilde{[W]}\), and \(\widetilde{[W^*]}\), which leads to discrepancies in the initial values when compared to the full model. In the QSSA model, \({d\widetilde{[WE_1]}}/{dt} = 0\) according to \eqref{PdPC_QSSA}, which results in a constant \(\widetilde{[WE_1]}\) (red dashed line in Fig. \ref{Fig_PdPC_QSSA}(b)). This, in turn, introduces an initial error in \(I_{\mathrm{ATP}}\). In our setup, since \(q_1 \ll a_1\), \(I_{\mathrm{ATP}}\) in the QSSA model can be approximated as \(k_1\widetilde{[WE_1]}\), making \(I_{\mathrm{ATP}}\) effectively constant (red dashed line in Fig. \ref{Fig_PdPC_QSSA}(e)). Similarly, \(I_{\mathrm{ADP}}\) is approximately constant and equals to \(-k_1\widetilde{[WE_1]}\) (yellow dashed line in Fig. \ref{Fig_PdPC_QSSA}(e)). We observe that \(I_{\mathrm{ATP}}\) is predominantly positive during the reaction, except for an initial transient negative phase, whereas \(I_{\mathrm{ADP}}\) and \(I_{\mathrm{Pi}}\) remain consistently negative. This suggests that energy ($\mathrm{ATP}$) is ultimately consumed to drive the reaction, aligning with the biochemical context of the system.

\subsection{Sensitivity of the PdPC Model}

\begin{figure}[htbp]
\centering
 \includegraphics[width=0.9\linewidth]{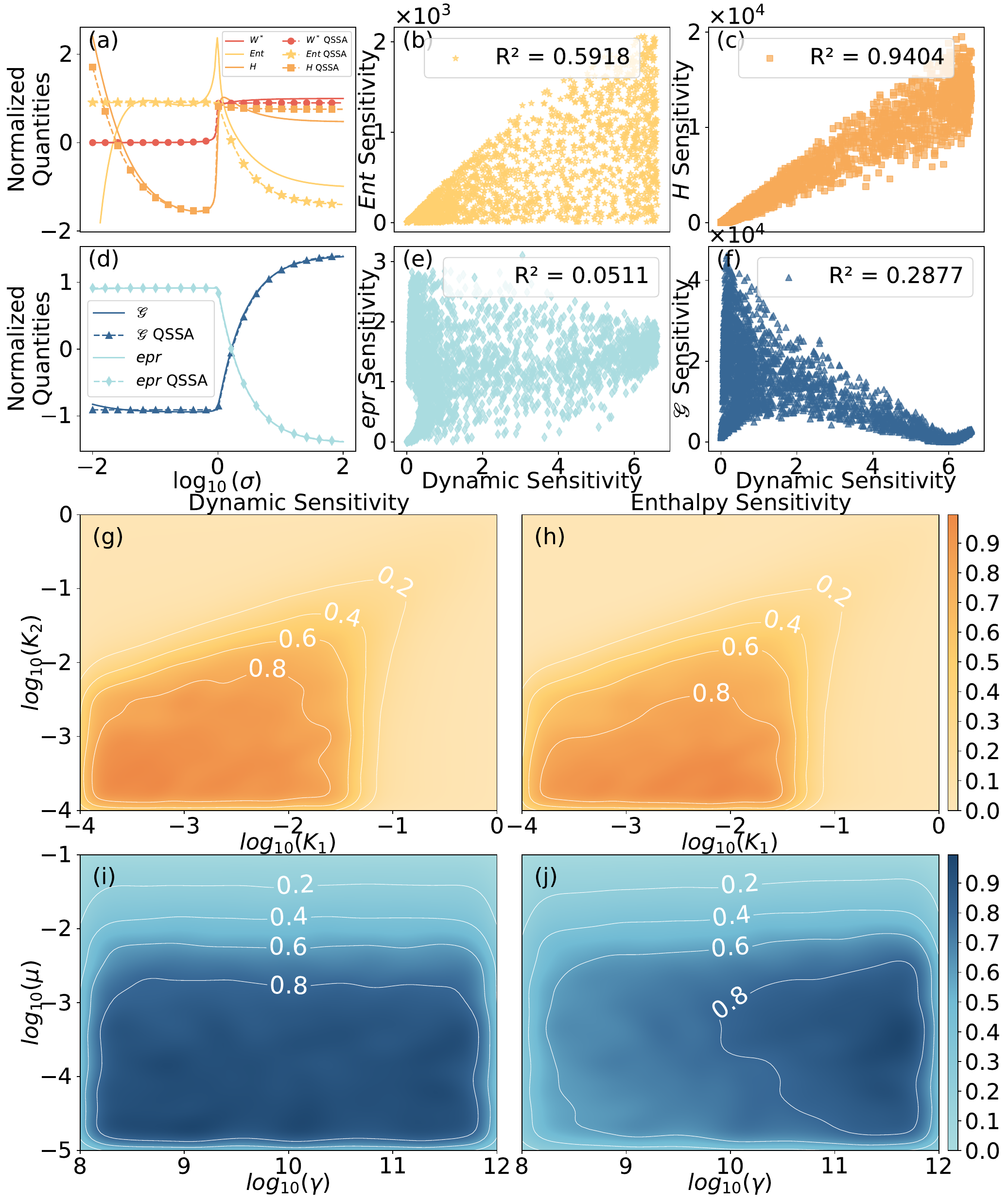}
\caption{\textbf{Comparison of Dynamic and Thermodynamic Sensitivities in the PdPC System.}
(a, d) \(W^*\) and steady-state thermodynamic quantities (solid lines), along with their QSSA-reduced counterparts (dashed lines with markers), plotted against \(\log_{10} \sigma\).
(b, c, e, f) Scatter plots of dynamic sensitivity versus thermodynamic sensitivities defined based on entropy (\(R^2 = 0.5918\)), enthalpy (\(R^2 = 0.9404\)), entropy production rate (\(R^2 = 0.0511\)), and intrinsic Gibbs free energy (\(R^2 = 0.2877\)) respectively. 
(g–j) Dynamic and enthalpy sensitivity across different parameter regions projected onto the \(\log_{10} K_1\)–\(\log_{10} K_2\) (orange) and \(\log_{10} \mu\)–\(\log_{10} \gamma\) (blue) planes.
}
\label{fig:Switchability}
\end{figure}

The \emph{switchability} of the PdPC system refers to the sigmoidal response of the phosphorylated protein fraction \( W^* = [W^*]/[W]_{\text{tot}} \) to variations in the composite parameter \( \sigma = (\kappa_2^+ [E_{1}]_{tot})/(\kappa_4^+ [E_{2}]_{tot}) \)\cite{Qian2003}, as illustrated by the red curve in Fig.~\ref{fig:Switchability}(a). 
To quantify how sharply the steady-state output \( W^* \) responds to perturbations in \(\sigma\) near the transition point, the  \emph{dynamic sensitivity} is defined as the absolute slope of \( W^* \) with respect to \(\sigma\) at \(\sigma = 1\): 
\begin{equation}
\mathcal{S}_{\text{dyn}} = \left|\frac{dW^*}{d\sigma}\right|_{\sigma=1}.
\end{equation}

In parallel, we define \emph{thermodynamic sensitivity} of the PdPC system as the absolute slope of a relevant steady-state thermodynamic quantity with respect to $\sigma$ at \(\sigma = 1\), which captures how sensitively the thermodynamic state of the system responds to external perturbations. 
To systematically analyze the thermodynamic and dynamic sensitivities, the thermodynamic quantities in Figs. \ref{fig:Switchability}(a, d) are linearly normalized. For example, the normalized enthalpy is given by $H_{\text{norm}}(\alpha) = (H(\alpha) - \bar{H})/{\operatorname{std}(H)}$, here \(\bar{H} = \sum_{\alpha} H(\alpha)/N\) is the mean and \(\operatorname{std}(H) = \sqrt{\sum_{\alpha} (H(\alpha) - \bar{H})^2 / N}\) is the standard deviation on all samples \(\alpha\). 

The thermodynamic quantity–\(\sigma\) curves are shown in Figs. \ref{fig:Switchability}(a, d). 
The entropy $Ent$, enthalpy $H$, intrinsic Gibbs free energy $\mathcal{G}$, and entropy production rate $epr$ all exhibit sharp transitions near \(\sigma = 1\), analogous to the behavior of \( W^* \). {This indicates that \(\sigma = 1\) serves as a critical transition point, at which the behaviors of both dynamic and thermodynamic quantities make changes not only in quantity, but also in quality.}

As shown through scatter plots in Figs. \ref{fig:Switchability}(b, c, e, f), the linear correlations between dynamic sensitivity and different thermodynamic sensitivities are examined under various parameter combinations. In particular, a near-perfect linear relationship between the dynamic sensitivity and the enthalpy sensitivity (\(R^2 = 0.9404\)) is found, suggesting a remarkably strong alignment between these two types of sensitivity.

Motivated by this, we first randomly perturb the four hyperparameters \(K_1=(\kappa_1^-+\kappa_2^+)/(\kappa_1^+[W]_{tot})\), \(K_2=(\kappa_3^-+\kappa_4^+)/(\kappa_3^+[W]_{tot})\), \(\mu=(\kappa_3^-\kappa_4^-)/(\kappa_3^+\kappa_4^+)\), and \(\gamma=(\kappa_1^+\kappa_2^+\kappa_3^+\kappa_4^+)/(\kappa_1^-\kappa_2^-\kappa_3^-\kappa_4^-)\), which, aside from \(\sigma\), uniquely determine the steady state of the system\cite{Qian2003}. Then we compute the corresponding pairs of dynamic and thermodynamic sensitivities, and project them onto the hyperparameter planes, \(K_1\)–\(K_2\) (orange) and \(\mu\)–\(\gamma\) (blue), as shown in Figs. \ref{fig:Switchability}(g–j). Prior to projection, both sensitivity values are rescaled to the same numerical range \((0,1)\) for direct comparison. 
Figs. \ref{fig:Switchability}(g-j) show the correlation between dynamic and thermodynamic sensitivities across different parameter regimes, revealing the extent to which the PdPC's dynamic responsiveness is coupled with the underlying thermodynamic driving forces. 
In both projections, enthalpy and dynamic sensitivities exhibit nearly identical spatial patterns, underscoring their high concordance. 
These results suggest that the enthalpy sensitivity can serve as a remarkably accurate predictor of dynamic sensitivity. Our finding marks a significant step toward the thermodynamics-assisted understanding and prediction of kinetic behavior in biochemical networks. 

\section{Conclusions}
In this work, we have established the effective nonequilibrium thermodynamics for open CRNs reduced by PEA and QSSA. 
The quantitative connection between the original mass-action equations and the PEA or QSSA-reduced models in the thermodynamic quantities, including the enthalpy, entropy production rate, free energy dissipation rate, and intrinsic Gibbs free energy, have been revealed and summarized via Eqs. \eqref{enthalpy-PEA}-\eqref{Qhk QSSA} and Table \ref{table1}. More importantly, several different versions of the second law of thermodynamics have been discussed. We demonstrate that the entropy production rate rigorously preserves non-negativity under PEA, whereas the non-negativity of the free-energy dissipation rate and housekeeping heat, along with the monotonicity of the intrinsic Gibbs free energy, cannot be guaranteed due to the violation of mass-action laws. A similar conclusion holds for QSSA, in which, however, the intrinsic Gibbs free energy retains its monotonicity. The above general results have been carefully validated through the application to MM reactions and PdPC reactions as two concrete examples, both theoretically and numerically. The MM reactions demonstrate clearly the thermodynamic structures of the reduced mechanisms by PEA and QSSA; while the PdPC reactions provide an example for further discussions of biological signal transduction. The sensitivity of PdPC reactions is maintained, while the correlation between sensitivity and several thermodynamic quantities is partly destroyed by the adoption of QSSA. This demonstrates the thermodynamic complexity associated with QSSA, implying further explorations along this direction. 

With regard to potential applications of the current study, we are facing with CRNs with few observables instead of a full knowledge of CRNs in many real-world scenarios. The extension of the nonequilibrium thermodynamics of general CRNs to these cases remains challenging. Fortunately, our current study on the thermodynamics of reduced models may shed new light on this area. We leave this point to future research.

\section*{Appendix}
\counterwithout{equation}{section}  
\setcounter{equation}{0}            
\renewcommand{\theequation}{A\arabic{equation}}  




\subsection{Counterexamples of the Thermodynamic Quantities for Reduced Models}
\label{Counterexamples of the Thermodynamic Quantities for Reduced Models} 
\begin{figure}[ht]
    \centering
    \includegraphics[width=0.8\linewidth]{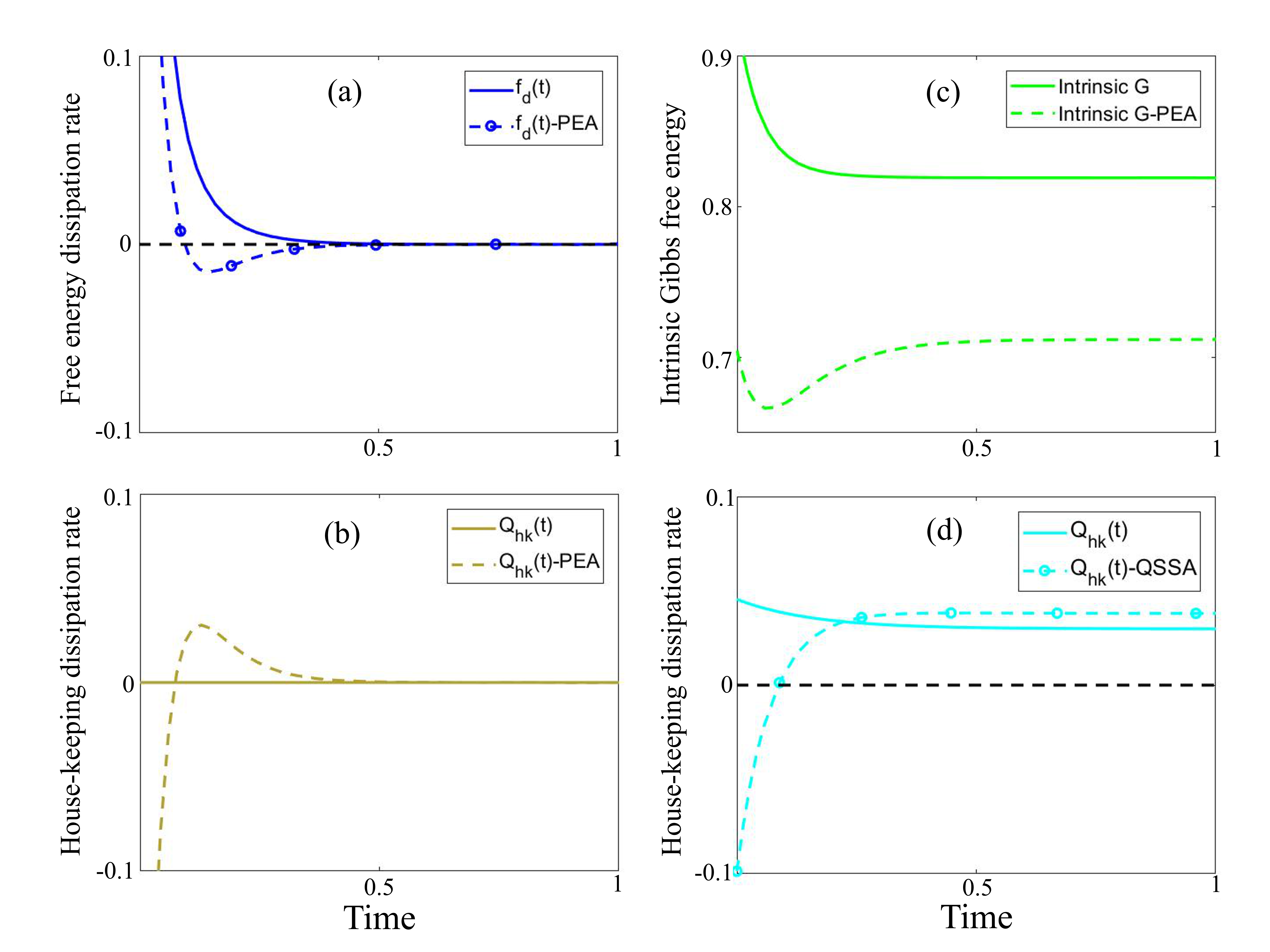}
    \caption{Illustration on the thermodynamic quantities for reduced models by (a,b,c) PEA and (d) QSSA. Trajectories of the (a) free energy dissipation rate, (b) house-keeping dissipation rate, and (c) intrinsic Gibbs free energy are shown for reduced models by PEA. Trajectory of the (d) house-keeping dissipation rate is shown for reduced models by QSSA. 
    The initial values and rate constants are taken as (a) $([S]_0, [ES]_0, [EP]_0; \kappa_1^+, \kappa_1^-, \kappa_2^+, \kappa_2^-, \kappa_3^+, \kappa_3^-)$ $=(0.5, 0.5, 0.5; 5,7.5,1.5,1.05,3.5,3.5/1.05)$ for (a); and $=(0.5, 0.5, 0.5; 9/2.5,9,2.5,1.5,3,2))$ for (b); $=(0.5, 0.5, 0.5; 6/2.5,6,2.5,1.5,3,2)$ for (c); $=(1,1,0.9; 2.5,0.8,1,3,3,4)$ for (d) correspondingly. We set the chemostatted species $[S]\equiv[P]\equiv 1$, the standard entropy of formation $\bm s^{\circ}=\bm 0$, the standard enthalpy of formation $\bm h^{\circ}$ based on the local detailed balance condition, and the  constants $\mathcal{R}=\mathcal{T}=1$.}
    \label{counterexample}
\end{figure}

As we have shown in mathematics that the reduced model by PEA or QSSA does not necessarily guarantee all fundamental thermodynamic properties of the full model. Here we make a numerical illustration through the open MM reactions. To be concrete, the negative part of the free energy dissipation rate for the PEA model is shown in Fig. \ref{counterexample} (a). 
The negative parts of the house-keeping dissipation rates for both the PEA and QSSA models are drawn in Fig. \ref{counterexample} (b) and (d). 
Specifically, the PEA-reduced model loses monotonicity in the intrinsic Gibbs free energy in Fig. \ref{counterexample} (c), whereas QSSA retains this property.

\subsection{Derivation of Thermodynamic Quantities in the Context of Ref. \cite{Avanzini2020}} 
\label{Derivation of the thermodynamic quantities in the framework} 
We proceed to derive the entropy production rate for the QSSA-reduced reactions in the context of Ref. \cite{Avanzini2020}. To avoid ambiguity, we retain the notation established in the main text to describe the following results of the reduced model based on Ref. \cite{Avanzini2020}. Recalling the effective stoichiometric matrix $\bm{\widetilde \nu}^S$, and dividing $\bm{\widetilde \nu}^S$ into $\bm{\widetilde \nu}^S_1=(-1,1,-1,0)^{\dag}$ and $\bm{\widetilde \nu}^S_2=(0,0,-1,1)^{\dag}$, we have 
\begin{align}
\label{thermo_Avan}
\mathcal{T}\widetilde{epr}(t)
&=-\widetilde{\mu} \cdot \bm{\widetilde \nu}^S_1\psi _{1}-\widetilde{\mu}  \cdot \bm{\widetilde \nu}^S_2\psi _{2} \nonumber \\
&=(\mu_{E}+\mu_{S}-\mu_{EP})((\kappa_{2}^{-}[P]+\frac{\kappa_{1}^{+}\kappa_{3}^{+}}{\kappa_{1}^{-} + \kappa_{3}^{+}}[S])[E]-(\kappa_{2}^{+}+\frac{\kappa_{1}^{-}\kappa_{3}^{-}}{\kappa_{1}^{-} + \kappa_{3}^{+}})[EP])\\
&+(\mu_{S}-\mu_{P})(\kappa_{2}^{+}[EP] - \kappa_{2}^{-}[E][P]) \nonumber,
\end{align}
where $\widetilde{\mu}=(\mu_{E}, \mu_{EP}, \mu_{S}, \mu_{P})^{\dag}.$ Direct calculations verify that the two expressions of entropy production rates in Eqs. \eqref{thermo_QSSA} and \eqref{thermo_Avan} coincide according to the local detailed balance condition, $\mu^{\circ}_{E}+\mu^{\circ}_{S}-\mu^{\circ}_{EP}=\mathcal{RT} \ln((\kappa_{1}^{+} \kappa_{3}^{+})/(\kappa_{1}^{-} \kappa_{3}^{-}))$ and $\mu^{\circ}_{S}-\mu^{\circ}_{P}=\mathcal{RT} \ln((\kappa_{1}^{+} \kappa_{2}^{+} \kappa_{3}^{+})/(\kappa_{1}^{-} \kappa_{2}^{-} \kappa_{3}^{-}))$.  

In order to describe open CRNs, Ref.\cite{Avanzini2020} introduced the concept of ``moiety”, which represents the part (or entire) of molecules that is exchanged with the environment through the chemostats. For the elementary dynamics, their concentration vector is specified as 
\begin{equation}
\bm{m}(\bm{c}(t))=\bm{M}^{-1}\bm{L}_{br}(\bm{c}(t)),
\end{equation}
while for the effective dynamics it is given by
\begin{equation}
\bm{\widetilde{m}}(\bm{c}^{S}(t))=\bm{M}^{-1}\bm{\widetilde{L}}_{br}(\bm{c}^{S}(t)),
\end{equation}
where the matrix \(\bm{M} = [\left(M_{ij}\right)]_{B \times B}\), with entries corresponding to the chemostatted species that breaking the conservation laws, \(M_{ij} = b_{ij}~(j \in Y_P)\). Here \(Y_P\) denotes the index of the columns for the chemostatted species that breaking the conservation laws. For example, for the MM reactions, the broken conservation law is $\bm{\ell}^{\zeta_{b}} $=$\bm{\ell}_{2}$=$(1,0,1,1,1)^{\dag}$, and $\bm{Y}_p=P$, we thus have $\bm{M}=(1)_{1\times 1}$. Therefore, for the full MM reactions, ${L}_{br}(\bm{c})$=(\dots, $\bm{\ell}^{\zeta_{b}}$ $\bm{c}$, \dots )=$[ES]+[EP]+[S]+[P]$, we have ${m}(\bm{c})=[ES]+[EP]+[S]+[P]$. Meanwhile, for the QSSA-reduced MM reactions, ${\widetilde{L}}_{br}(\bm{c}^{S})=[EP]+[S]+[P]$, we have ${\widetilde{m}}(\bm{c}^{S})=[EP]+[S]+[P]$.

From the above description, the semigrand Gibbs free energy $\mathscr{G}(t)$ for the full MM reactions is derived by eliminating the energetic contributions of the matter exchanged with the reservoirs \cite{Avanzini2020}. That is, 
\begin{align*}
\mathscr{G}(t)
&=G(t)-\bm{\mu}_{Y_{P}}(\bm{c})\cdot\bm{m}(\bm{c})\\
&={\mu}_{ES}[ES]+{\mu}_{E}[E]+{\mu}_{EP}[EP]+{\mu}_{S}[S]-\mathcal{RT}([ES]+[E]+[EP]+[S+[P])
-{\mu}_{P}(ES]+[EP]+[S]).
\end{align*}
Correspondingly, the semigrand Gibbs free energy for the QSSA-reduced MM reactions becomes 
\begin{align*}
\widetilde{\mathscr{G}}(t)
&= \widetilde{G}(t)-\bm{\mu}_{Y_{P}}(\bm{c}^{S})\cdot\widetilde{\bm{m}}(\bm{c}^{S})\\
&={\mu}_{E}[E]+{\mu}_{EP}[EP]+{\mu}_{S}[S]+{\mu}_{P}[P]-\mathcal{RT}([E]+[EP]+[S+[P])-{\mu}_{P}([EP]+[S]+[P])\\
&={\mu}_{E}[E]+{\mu}_{EP}[EP]+{\mu}_{S}[S] - \mathcal{RT}([E]+[EP]+[S+[P])-{\mu}_{P}([EP]+[S]).
\end{align*}

Let us explain the properties of $\widetilde{\mathscr{G}}(t)$ in the following.
The time evolution of $\widetilde{\mathscr{G}}(t)$ is separated into three parts, consisting of the entropy production rate $-\mathcal{T} \widetilde{epr}(t)$, the deriving work rate $\widetilde{ \overset{.}{\omega }}_{driv}(t)$, and the non-conservative work rate $\widetilde{ \overset{.}{\omega }}_{nc}(t)$: 
\begin{equation}
d_{t}\widetilde{\mathscr{G}}(t)=-\mathcal{T} \widetilde{epr}(t)+\widetilde{ \overset{.}{\omega }} _{driv}(t)+\widetilde{ \overset{.}{\omega }}_{nc}(t).
\end{equation}
Specifically speaking, 
\begin{align*}
d_{t}\widetilde{\mathscr{G}}(t)
&=(\mu_{E},\mu _{EP},\mu _{S},\mu _{P})\cdot
\frac{d}{dt}([E],[EP],[S],[P])^{\dag}
+\frac{d}{dt}(\mu _{E},\mu _{EP},\mu_{S},\mu _{P})\cdot([E],[EP],[S],[P])^{\dag}
\\&-{\mu}_{P}\frac{d}{dt}([EP]+[S]+[P])-([EP]+[S]+[P])\frac{d{\mu}_{P}}{dt}
-\mathcal{RT} \frac{d}{dt}([E]+[EP]+[S]+[P]),
\end{align*}
where 
\begin{align*}
&\frac{d}{dt}(\mu _{E},\mu _{EP},\mu_{S},\mu _{P})\cdot([E],[EP],[S],[P])^{\dag}\\
&=\mathcal{RT}(\frac{1}{[E]}\frac{d[E]}{dt}, \frac{1}{[EP]}\frac{d[EP]}{dt}, \frac{1}{[S]}\frac{d[S]}{dt}, \frac{1}{[P]}\frac{d[P]}{dt})\cdot([E],[EP],[S],[P])^{\dag}\\
&=\mathcal{RT}\frac{d}{dt}([E]+[EP]+[S]+[P]),
\end{align*}
and
\begin{equation*}
-\mu_{p} \frac{d}{dt}([EP]+[S]+[P])=-\mu_{p}(I^{S} +I^{P} ).
\end{equation*}

Therefore,
\begin{align*}
d_{t}\widetilde{\mathscr{G}}(t)
&=(\mu _{E},\mu _{EP},\mu _{S},\mu _{P})[\widetilde{\bm\nu}^{S}(\psi _{1},\psi _{2})^{\dag}
+(0, 0, I^{S}, I^{P})^{\dag}] 
- \frac{d\mu _{P}}{dt} {\widetilde{m}}(\bm{c}^{s})-\mu_{P}(I^{S} +I^{P} )\\
&=(\mu _{E},\mu _{EP},\mu _{S},\mu _{P})^{\dag}
\widetilde{\bm \nu}^{S}(\psi _{1},\psi _{2})^{\dag}
- \frac{d\mu_{P}}{dt} {\widetilde{m}}(\bm{c}^{s}) + I^{S}(\mu_{S} -\mu _{P})\\
&=-\mathcal{T}\widetilde{epr}(t)+\widetilde{ \overset{.}{\omega }} _{driv}(t)+\widetilde{ \overset{.}{\omega }}_{nc}(t), 
\end{align*}
where the deriving work rate is
\begin{equation*}
\widetilde{ \overset{.}{\omega }}_{driv}(t)
=-\frac{d{\mu}_{P}}{dt} {\widetilde{m}}(\bm{c}^{s})
=- \frac{d{\mu}_{P}}{dt} ([EP]+[S]+[P]),
\end{equation*}
and the nonconservative work rate is 
\begin{equation*}
\widetilde{ \overset{.}{\omega }}_{nc}(t)
=I^{S}(\mu_{S} -\mu_{P}).
\end{equation*}
We conclude that the semigrand Gibbs free energy defined in \cite{Avanzini2020} is fundamentally distinct from the intrinsic Gibbs free energy introduced in this work. Crucially, this distinction persists even under the condition ${d{\mu}_{P}}/{dt}=0$, as the non-conservative work rate $\widetilde{ \overset{.}{\omega }}_{nc}(t) \neq 0$.

\section*{Data availability statement}
The data cannot be made publicly available upon publication because they are not available in a format that is sufficiently accessible or reusable by other researchers. The data that support the findings of this study are available upon reasonable request from the authors.
\section*{Acknowledgments}
The authors acknowledged the financial supports from the National Key R\&D Program of China (Grant No. 2023YFC2308702), the National Natural Science Foundation of China (Grants No. 12205135), the Natural Science Foundation of Fujian Province of China (2024J01212), Guangdong Basic and Applied Basic Research Foundation (2023A1515010157). 
The authors thank Dr. Wuyue Yang for valuable discussions.
\bibliographystyle{unsrt}%
\bibliography{fp}
\end{document}